\RequirePackage{fix-cm}
\documentclass[smallcondensed,natbib]{svjour3}

\smartqed  
\usepackage[utf8]{inputenc}
\usepackage{graphicx}   
\usepackage{multirow}   
\usepackage{diagbox}   
\usepackage{makecell}   
\usepackage{amsmath}     
\usepackage{tcolorbox}  
\usepackage{xcolor,colortbl,soul}  
\usepackage{tikz}       
\usetikzlibrary{shapes.arrows}
\usepackage[breaklinks]{hyperref}	
\usepackage[all]{hypcap}            

\definecolor{DarkOrange}{RGB}{255,80,0}

\hypersetup{
    colorlinks=true,
    linkcolor=blue,
    citecolor=blue,
    urlcolor=blue,
}

\usepackage{etoolbox}
\makeatletter
\patchcmd{\NAT@citex}
  {\@citea\NAT@hyper@{%
     \NAT@nmfmt{\NAT@nm}%
     \hyper@natlinkbreak{\NAT@aysep\NAT@spacechar}{\@citeb\@extra@b@citeb}%
     \NAT@date}}
  {\@citea\NAT@nmfmt{\NAT@nm}%
   \NAT@aysep\NAT@spacechar\NAT@hyper@{\NAT@date}}{}{}
\patchcmd{\NAT@citex}
  {\@citea\NAT@hyper@{%
     \NAT@nmfmt{\NAT@nm}%
     \hyper@natlinkbreak{\NAT@spacechar\NAT@@open\if*#1*\else#1\NAT@spacechar\fi}%
       {\@citeb\@extra@b@citeb}%
     \NAT@date}}
  {\@citea\NAT@nmfmt{\NAT@nm}%
   \NAT@spacechar\NAT@@open\if*#1*\else#1\NAT@spacechar\fi\NAT@hyper@{\NAT@date}}
  {}{}
\makeatother

\def\checkmark{\tikz\fill[scale=0.4](0,.35) -- (.25,0) -- (1,.7) -- (.25,.15) -- cycle;} 
\newcommand{\UpArrow}{\begin{tikzpicture}[baseline=-0.3em]
\node[single arrow,draw,rotate=90,single arrow head extend=0.2em,inner
ysep=0.2em,transform shape,line width=0.05em,fill=green] (X){};
\end{tikzpicture}}  
\newcommand{\Dash}{\begin{tikzpicture}[baseline=-0.35em]
\draw[line width=0.2em] (0,0) -- (0.33,0);
\end{tikzpicture}}  
\newcommand{\DownArrow}{\begin{tikzpicture}[baseline=-0.4em]
\node[single arrow,draw,rotate=270,single arrow head extend=0.2em,inner
ysep=0.2em,transform shape,line width=0.05em,fill=red] (X){};
\end{tikzpicture}}  
\def\checkmark{\tikz\fill[scale=0.4](0,.35) -- (.25,0) -- (1,.7) -- (.25,.15) -- cycle;}

\begin{document}

\title{An Empirical Study of Developers’ Discussions about Security Challenges of Different Programming Languages}

\author{Roland Croft \and
        Yongzheng Xie \and
        Mansooreh Zahedi \and
        M. Ali Babar \and
        Christoph Treude
}

\institute{Roland Croft, Yongzheng Xie, Ali Babar \at
            School of Computer Science, The University of Adelaide, Australia \\
            \email{firstname.lastname@adelaide.edu.au}
            \and
            Mansooreh Zahedi, Christoph Treude \at
            School of Computing \& Information Systems, The University of Melbourne, Australia \\
            \email{firstname.lastname@unimelb.edu.au}
            \and
            Roland Croft, Ali Babar \at
            Cyber Security Cooperative Research Centre
}

\date{Received: date / Accepted: date}
\maketitle

\begin{abstract}
Given programming languages can provide different types and levels of security support, it is critically important to consider security aspects while selecting programming languages for developing software systems. Inadequate consideration of security in the choice of a programming language may lead to potential ramifications for secure development. Whilst theoretical analysis of the supposed security properties of different programming languages has been conducted, there has been relatively little effort to empirically explore the actual security challenges experienced by developers. We have performed a large-scale study of the security challenges of 15 programming languages by quantitatively and qualitatively analysing the developers’ discussions from Stack Overflow and GitHub. By leveraging topic modelling, we have derived a taxonomy of 18  major security challenges for 6 topic categories. We have also conducted comparative analysis to understand how the identified challenges vary regarding the different programming languages and data sources. Our findings suggest that the challenges and their characteristics differ substantially for different programming languages and data sources, i.e., Stack Overflow and GitHub. The findings provide evidence-based insights and understanding of security challenges related to different programming languages to software professionals (i.e., practitioners or researchers). The reported taxonomy of security challenges can assist both practitioners and researchers in better understanding and traversing the secure development landscape. This study highlights the importance of the choice of technology, e.g., programming language, in secure software engineering. Hence, the findings are expected to motivate practitioners to consider the potential impact of the choice of programming languages on software security.
\end{abstract}

\section{Introduction}
A recent successful cybersecurity attack on FireEye\footnote{\url{https://www.nytimes.com/2020/12/08/technology/fireeye-hacked-russians.html}}, one of the World’s largest security companies, is yet another indicator of the rapidly increasing security threats to digital systems \citep{rafter2019}, which are enabled by software. To ensure the availability, confidentiality and integrity of a software system, it is vital that organisations resolve software security issues that may be introduced during development and/or evolution of a software system \citep{shahriar2012}. That is why there are increasing calls for putting more emphasis on software security issues throughout software development and evolution by incorporating the principles, practices, and tools of software security \citep{mailloux2018}. There is an important need to provide developers with valuable security information sources and support to achieve secure software development; most novice developers lack the required security expertise \citep{barnum2005}. 


Whilst developers play a key role in the choice of an implementation technology, e.g., programming language, for a software development project, developers may tend to consider the choice of a programming language from the perspectives of project requirements, personal experience, learning costs, language popularity, and the availability of toolkits/packages \citep{meyerovich13}. Given that development security flaws and weaknesses can cause major software security problems, it is important to consider the security issues when choosing programming languages, which are known to provide different types and levels of support for security \citep{shahriar2012}. For example, the C or C++ programming languages can have un-trapped memory management errors that make them unsafe; however, the Java programming language provides static and dynamic checks that ensure no un-trapped memory management errors can occur. We assert that it is important to increase awareness and knowledge about the security issues encountered while using different programming languages. Our research, reported in this paper, purports to empirically identify the security challenges that developers experience while using different programming languages. We decided to analyse and synthesize software developers’ observations and experiences of security issues related to different programming languages shared on Open Data platforms like Stack Overflow and GitHub.

To achieve this objective, we have conducted a large-scale study of developers’ discussions about security issues related to popular programming languages using quantitative and qualitative methods. For this study, we examine over 20 million issues from 27,312 GitHub repositories and over 11 million posts from Stack Overflow, for 50 of the most popular programming languages. We have then conducted an in-depth analysis of 136,400 issues from 7194 GitHub repositories and 143,498 Stack Overflow posts  for 15 of the most popular programming languages such as Java, C, Python, and PHP. 

We assert that this study has gathered and analysed the largest amount of security issue related data generated from developers’ discussions on Stack Overflow and GitHub. This study also provides useful insights into three major software engineering aspects: the projects, the problems and the users. Whilst a few efforts have performed exploratory analysis of the security topics discussed by developers on Stack Overflow \citep{yang2016} and GitHub \citep{zahedi2018}, there has been no effort aimed at empirically exploring the security issues from a technology or programming language perspective. Furthermore, we are the first to conduct comparative analysis of the experienced security challenges across data sources and domains. Previous studies have similarly investigated the characteristics of Stack Overflow security discussions \citep{bayati2016,lopez2018,lopez2019}, to better understand the nature and ways in which security knowledge is shared. We extend this analysis by comparing the characteristics of specific security topics and domains to investigate the explicit differences in the challenges by examining the discussion popularity, difficulty, and expertise. The main findings from this study are:

\begin{itemize}
    \item Overall, the security challenges are increasing for most languages, except for PHP. Newer languages (i.e., Swift and TypeScript) experience a constant rise in security challenges after release.
    \item Security challenges and trends often cluster based on the tasks that the languages are commonly used for. GitHub security issues of Web-oriented languages (e.g., JavaScript, TypeScript and PHP) receive the most popularity, and GitHub users for Mobile-oriented languages (e.g., Java, C\#, Objective-C and Swift) have the highest level of security expertise.
    \item Shell-based and Web-oriented languages experience significantly higher average rates of security discussion. Julia, MATLAB and R (languages primarily for scientific programming) exhibit exceedingly small numbers of security discussion. 
    \item C/C++ are the only languages which prominently face challenges for memory management, and their questions on Stack Overflow receive some of the most popularity and attention from expert users. C/C++ users also have the most consideration for security via the intent of their discussions.  
    \item There is a disconnect in the nature of the language security challenges between Stack Overflow and GitHub. In particular, Web-Development languages and Authentication challenges are considered easier and less popular on Stack Overflow than they are on GitHub.
\end{itemize}

The findings from this study are expected to provide better understanding and useful insights about the security issues experienced when using different programming languages. Software development practitioners can gain better understanding of the influence that programming languages have on software security. Developers can be motivated to consider the security issues associated with different programming languages when choosing and using different programming languages in software development projects. The main contributions of our work reported in this paper are as follows:

\begin{itemize}
    \item It carries out a first of its kind empirical study of security issues for different programming languages by quantitatively and qualitatively analysing the largest amount of data of developers’ discussions on Stack Overflow and GitHub. The findings contribute to the evidence-based body of knowledge about the security issues related to programming languages.
    \item It identifies a taxonomy of 18 security topics and 6 security categories that are commonly discussed and faced by developers during software development. The identified taxonomy can help categorize known security issues related to different programming languages as a source of guidance for developers. 
    \item It provides a method that enabled us to extract, cluster, and analyse a large number of domain specific discussions from both Stack Overflow and GitHub. The reported method helped us to construct and make publicly available the largest, to our knowledge, dataset of developers’ discussions about security issues of 15 popular programming languages\footnote{\url{https://github.com/RolandCroft/Language-Security-Challenges/}}.
\end{itemize}

The rest of the paper is organized as follows. In Section 2, we discuss the background and motivation of this study. In Section 3, we present the methodological and logistical details for conducting this research. We report and analyse the results in Section 4. We discuss the results and highlight the implications of this research for practitioners and researchers in Section 5. The threats to validity and their implications are presented in Section 6. Section 7 draws some conclusions and indicates the possible future work. 

\section{Background and Motivations}
\subsection{Programming Language Software Quality Analysis}
Programming language paradigms define the characteristics of a programming language and provide classification \citep{pierce2002}. In Table \ref{table:paradigms}, we categorize 15 popular programming languages using a classification system inline with previous works \citep{ray2014, zhang2019}. We note that there are no commonly agreed definitions for language types. It is also known that paradigms are often relative rather than absolute.  We define the type safety as whether a language prevents operations between mismatched types (strong) or allows type inconsistency (weak) \citep{cardelli85}. Type checking determines whether type errors are reported based on the source code (static) or the run-time behaviour (dynamic) \citep{cardelli85}. Memory management refers to the extent at which the programs handle memory safety, control and management; whether the responsibility is on the user (manual) or not (managed) \citep{dhurjati2003}. 

\begin{table}[h]
\caption{Categories of 15 popular programming languages}
\label{table:paradigms}
\resizebox{\columnwidth}{!}{%
\begin{tabular}{ |p{2cm}|c|c|c|c|c|c|c|c|c|c|c|c|c|c|c| } 
    \hline
    \multicolumn{2}{|c|}{\diagbox{Category}{Language}} & Java & PHP & JavaScript & C\# & Python & Shell & C/C++ & Ruby & Objective-C & Swift & PowerShell & Perl & Go & TypeScript \\ 
    \hline
    \multirow{2}{2cm}{Type Safety} & Strong & \cellcolor{lightgray} \checkmark & & \cellcolor{lightgray} \checkmark & \cellcolor{lightgray} \checkmark & \cellcolor{lightgray} \checkmark & & & \cellcolor{lightgray} \checkmark & & \cellcolor{lightgray} \checkmark & \cellcolor{lightgray} \checkmark & \cellcolor{lightgray} \checkmark & \cellcolor{lightgray} \checkmark & \cellcolor{lightgray} \checkmark\\
     & Weak & & \cellcolor{lightgray} \checkmark & & & & \cellcolor{lightgray} \checkmark & \cellcolor{lightgray} \checkmark & & \cellcolor{lightgray} \checkmark & & & & & \\
    \hline
    \multirow{2}{2cm}{Type Checking} & Static & \cellcolor{lightgray} \checkmark & & & \cellcolor{lightgray} \checkmark & & & \cellcolor{lightgray} \checkmark & & & \cellcolor{lightgray} \checkmark & & & \cellcolor{lightgray} \checkmark & \\
     & Dynamic & & \cellcolor{lightgray} \checkmark & \cellcolor{lightgray} \checkmark & & \cellcolor{lightgray} \checkmark & \cellcolor{lightgray} \checkmark & & \cellcolor{lightgray} \checkmark & \cellcolor{lightgray} \checkmark & & \cellcolor{lightgray} \checkmark & \cellcolor{lightgray} \checkmark & & \cellcolor{lightgray} \checkmark\\
    \hline
    \multirow{2}{2cm}{Memory Management} & Managed & \cellcolor{lightgray} \checkmark & \cellcolor{lightgray} \checkmark & \cellcolor{lightgray} \checkmark & \cellcolor{lightgray} \checkmark & \cellcolor{lightgray} \checkmark & \cellcolor{lightgray} \checkmark & & \cellcolor{lightgray} \checkmark & & \cellcolor{lightgray} \checkmark & \cellcolor{lightgray} \checkmark & \cellcolor{lightgray} \checkmark & \cellcolor{lightgray} \checkmark & \cellcolor{lightgray} \checkmark \\
     & Manual & & & & & & & \cellcolor{lightgray} \checkmark & & \cellcolor{lightgray} \checkmark & & & & & \\
    \hline
    \multicolumn{2}{|l|}{Creation Date} & 1995 & 1995 & 1995 & 2000 & 1990 & 1989 & 1972 & 1995 & 1984 & 2014 & 2006 & 1987 & 2009 & 2012\\
    \hline
\end{tabular}%
}
\end{table}

Many studies have investigated various attributes and aspects of programming languages \citep{horschig2018, bhattacharya2011, kleinschmager2012, sestoft2005, hanenberg2014}, through empirical analysis of historical software artefacts. However, they fail to provide a developers' perspective of the actual experienced challenges. 

Only a few works aim to form a large-scale comparative analysis between programming languages. \cite{ray2014} investigated the impacts of programming language paradigms on various code quality metrics, \cite{kochhar2016} similarly investigated the effects of multiple programming languages on code quality, and \cite{zhang2019} analysed the impacts of programming language on bug resolution characteristics. We aim to build upon the previous works and provide a security perspective. 

Similar to how previous works indicate that choice of a programming language may impact code quality \citep{ray2014, kochhar2016, zhang2019}, we extrapolate that programming language selection will also impact code security. For instance, type safe languages (languages which check for type errors \citep{cardelli85}) are usually considered resistant to errors such as Buffer Overflow, as they handle typing and other memory management tasks automatically \citep{grossman2005}. In contrast, a developer is responsible for ensuring safety and protection in weak typed languages, which leaves them more open to possible vulnerabilities and exploits. Thus, it is expected that the programming paradigms would impact the relative security and safety of different programming languages. 

\cite{khwaja2020} identify the different security vulnerabilities present in different programming languages, and the languages' support for prevention. Similarly, \cite{cifuentes2019} perform a theoretical analysis of different languages capabilities to handle different vulnerabilities. However, they only consider a subset of vulnerabilities and their analysis remains purely theoretical; security flaws do not directly equate to development challenges. 

Furthermore, the security requirements and challenges of software are dependent on the purposes it is used for \citep{sindre2005}. We would expect that network security and access control are especially important for web development, whereas resource management errors are more important for systems programming. However, different programming languages are suited better to different development tasks and domains. We outline some common development areas for 15 popular programming languages in Table \ref{table:domains}, based on the common uses and professions of each language described in popular online blogs\footnote{\url{https://raygun.com/blog/programming-languages/}} \footnote{\url{https://www.computerscience.org/resources/computer-programming-languages/}}. Hence, we anticipate that the relative security requirements and considerations differ for each programming language based on the activities it is used for. 

\begin{table}[h]
\caption{Common development domains of 15 popular programming languages}
\label{table:domains}
\resizebox{\columnwidth}{!}{%
\begin{tabular}{ |c|c|c|c|c|c|c|c|c|c|c|c|c|c|c| } 
    \hline
    \diagbox{Development}{Language} & Java & PHP & JavaScript & C\# & Python & Shell & C/C++ & Ruby & Objective-C & Swift & PowerShell & Perl & Go & TypeScript \\ 
    \hline
    Front-End Web & & & \cellcolor{lightgray} \checkmark & & & & & & & & & & & \cellcolor{lightgray} \checkmark \\
    \hline
    Back-End Web & \cellcolor{lightgray} \checkmark & \cellcolor{lightgray} \checkmark & \cellcolor{lightgray} \checkmark & \cellcolor{lightgray} \checkmark & \cellcolor{lightgray} \checkmark & & & \cellcolor{lightgray} \checkmark & & & & \cellcolor{lightgray} \checkmark & \cellcolor{lightgray} \checkmark & \cellcolor{lightgray} \checkmark \\
    \hline
    Mobile & \cellcolor{lightgray} \checkmark & & \cellcolor{lightgray} \checkmark & \cellcolor{lightgray} \checkmark & & & & & \cellcolor{lightgray} \checkmark & \cellcolor{lightgray} \checkmark & & & & \\
    \hline
    Game & \cellcolor{lightgray} \checkmark & & & \cellcolor{lightgray} \checkmark & & & \cellcolor{lightgray} \checkmark & & & & & & & \cellcolor{lightgray} \checkmark \\
    \hline
    Application & \cellcolor{lightgray} \checkmark & & & \cellcolor{lightgray} \checkmark & \cellcolor{lightgray} \checkmark & & & & & \cellcolor{lightgray} \checkmark & & \cellcolor{lightgray} \checkmark & \cellcolor{lightgray} \checkmark & \\
    \hline
    Systems & & & & & & & \cellcolor{lightgray} \checkmark & & & & & & \cellcolor{lightgray} \checkmark & \\
    \hline 
    Data Science & \cellcolor{lightgray} \checkmark & & & & \cellcolor{lightgray} \checkmark & & \cellcolor{lightgray} \checkmark & & & & & & & \\
    \hline
    Cloud & \cellcolor{lightgray} \checkmark & \cellcolor{lightgray} \checkmark & \cellcolor{lightgray} \checkmark & \cellcolor{lightgray} \checkmark & & & & \cellcolor{lightgray} \checkmark & & & & & \cellcolor{lightgray} \checkmark & \\
    \hline
    Command-Line & & & & & & \cellcolor{lightgray} \checkmark & & & & & \cellcolor{lightgray} \checkmark & & & \\
    \hline
\end{tabular}%
}
\end{table}

However, the extent to which these security requirements and challenges differ for each programming language is presently unknown, and current knowledge is based on theory, personal opinion, or anecdotal evidence. Hence, we aim to investigate developer knowledge, views and experiences to help build an evidence-based body of knowledge about programming language related aspects of secure software engineering. 

\subsection{Publicly Available Security Discussions}
There exists an abundance of developer discussion and knowledge sharing in publicly available repositories. Question \& Answering (Q\&A) sites contain crowd-sourced knowledge on a variety of software development related tasks and subjects. Repository hosting sites document software development practices and issues. Thus, software researchers have started mining these data sources with the aim of understanding software practices and improving development processes. 

Thus, we also utilize these two sources to obtain a snapshot of security-related developer activities and knowledge. Due to the nature of these discussions, either asking for help or identifying an issue, we refer to the topic of a discussion as a \emph{challenge}. By obtaining knowledge of the security challenges that practitioners individually face for development in particular programming languages, we can form our comparative analysis of the manifested challenges to identify differences in their nature and characteristics. 

\begin{figure}[h]
\centering
    \includegraphics[height=0.4\textheight]{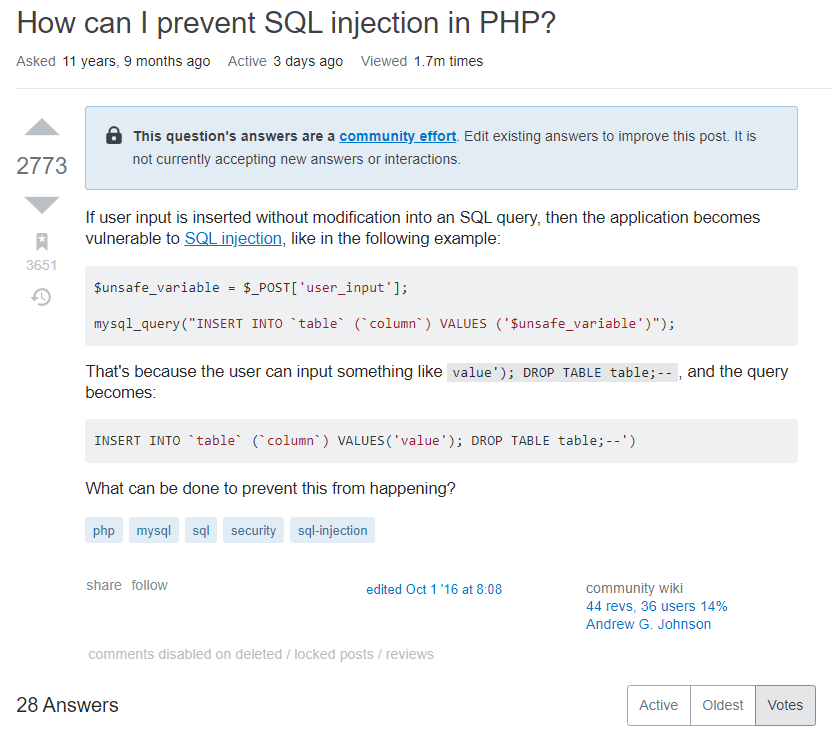}
    \caption{A Stack Overflow post about SQL injection prevention}
    \label{fig:so_example}
\end{figure}

\textbf{Stack Overflow} is a Q\&A website for software development. We select Stack Overflow for our data analysis as it is the most widely used source for Software Engineering (SE) based knowledge sharing. As of December 2020, there have been over 20 million questions and 30 million answers posted by over 13 million developers since the websites inception in 2008. Each Stack Overflow post contains a unique id, owner, title, question post, tag(s) and answer post(s) along with additional meta-information such as comments, views, favorites and score. For the purposes of this study, we henceforth use the term \textit{post} to refer to a question and all its respective answers. Example Stack Overflow posts are referenced in this paper via \textit{(SO, XXX)} which can be accessed through a web browser via https://stackoverflow.com/questions/XXX.  Fig. \ref{fig:so_example} shows an example Stack Overflow question about preventing the security vulnerability SQL injection. 

Stack Overflow is becoming of increasing focus to software researchers due to its diverse and thorough body of knowledge. In a survey conducted by \cite{ahmad2017} they discovered over 500 papers related to mining Stack Overflow data. \cite{le2020} construct a tool named PUMiner with the explicit purpose of mining security related documents from Q\&A sites like Stack Overflow. Whilst we do not explicitly adopt their tool, we derive heavy inspiration for the method of our data collection. 

\begin{figure}[h]
\centering
    \includegraphics[width=0.4\textheight]{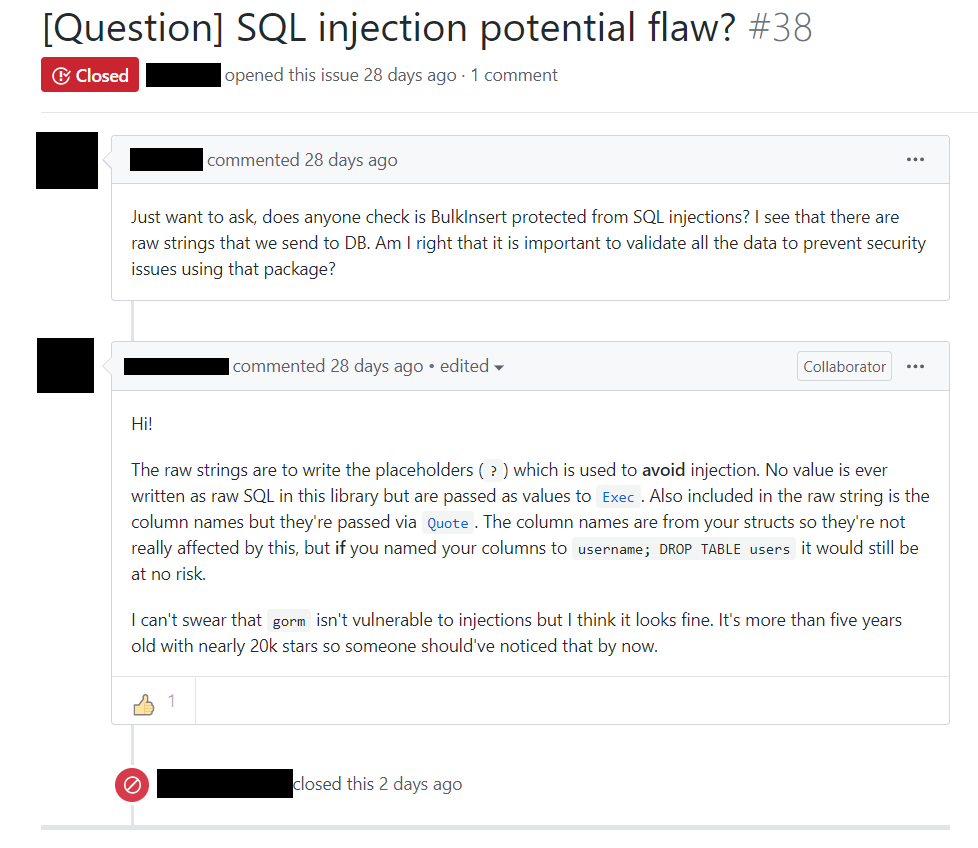}
    \caption{A GitHub issue about potential SQL injection}
    \label{fig:github_example}
\end{figure}

\textbf{GitHub} is an open-source platform for software repository hosting. We select GitHub as it is the most widely used public software development platform. As of December 2020, there are over 58 million registered users\footnote{\url{https://github.com/search?q=type:user\&type=Users}}, and over 41 million publicly hosted repositories\footnote{\url{https://github.com/search?q=is:public}}. Developers can stimulate discussion with other collaborators about ideas, enhancements, tasks or bugs via GitHub issues. We focus on GitHub issues to observe and analyse the problems and discussions that developers conduct during software development. An example GitHub issue discussing the possibility of an SQL Injection vulnerability in the software can be seen in Fig. \ref{fig:github_example}. GitHub repositories also provide metadata about their programming language composition, so we associate discussions with the main programming language of that repository. We use the term \textit{issue} to refer to a GitHub issue and all its associated comments. In this study, we reference example GitHub issues as \textit{(GH,XXX)} which can be accessed through a web browser via https://github.com/XXX. 

GitHub contains a much larger variety of data, and consequently its research applications are much more diverse. \cite{kalliamvakou2014} conduct an empirical analysis of GitHub data and warn researchers of perils of mining GitHub for research purposes. In particular, they find that the majority of repositories on GitHub are personal, inactive or not related to software development. We attempt to avoid these perils by only considering the most starred repositories. \cite{pletea2014} present sentiment analysis of security related discussions on GitHub. We build upon their method of data collection to extract security discussions from GitHub. 

There are also a few studies which aim to draw connections between Stack Overflow and GitHub data. \cite{vasilescu2013} investigate the activity correlation of users in Stack Overflow discussions and committers in GitHub repositories. They find that active GitHub users tend to ask fewer questions and provide more answers on Stack Overflow. A similar study is conducted by \cite{xiong2017} who create an algorithm for linking developer identity on Stack Overflow and GitHub, and analyse their behaviour on both sites. 

We observe Stack Overflow and GitHub to be two of the most prominent online sources for developer community support, due to their size and popularity. Hence, to obtain a more complete view of development security challenges, we intend to pull related insights from the discussions of both Stack Overflow and GitHub. 

\subsection{Topic Modelling of Software Engineering}
Topic modelling is a statistical Natural Language Processing (NLP) technique used to automatically discover the topics of text corpora \citep{blei2003}. Latent Dirichlet Allocation (LDA) is the most widely used implementation of topic modelling \citep{chen2016}, and hence we also adopt it as the standard. LDA uses word frequencies and co-occurrence frequencies in the input documents to build probabilistic word and document models that uncover the hidden semantic structures of the text \citep{blei2003}.

LDA has become a popular and effective tool for analysing large sets of software engineering related text data. Like us, many previous researchers have used LDA to automatically identify and cluster the topics of publicly accessible discussions on Stack Overflow for a variety of domains and aspects. \cite{barua2014}, \cite{zou2015}, and \cite{allamanis2013} have all leveraged LDA to identify general topics and trends in Stack Overflow data. This type of research has also been targeted towards a variety of different domains: mobile \citep{linares2013, rosen2016}, web development \citep{bajaj2014}, concurrency \citep{ahmed2018}, big data \citep{bagherzadeh2019}, API usage \citep{campbell2013}, machine learning \citep{bangash2019, han2020}, blockchain \citep{wan2019}, and security \citep{yang2016}. Some studies have also aimed to analyse the discussions of GitHub repositories like we do. \cite{rahman2014} apply LDA to repository pull requests in order to help identify factors that contribute towards successful or unsuccessful pull requests. \cite{zahedi2018} apply LDA to issue discussions in GitHub repositories to identify the main security issues that developers face. Due to its successful application to similar research domains as ours, we also utilize LDA to cluster the discussions and identify the main challenges. 

Whilst some of our topics overlap with the existing security-related LDA research \citep{yang2016,zahedi2018,le2020large}, our overall taxonomy differs substantially as we avoid technology specific topics and ensure generalizability for comparison. Furthermore, due to the consideration of programming language our analyzed posts differ substantially. Reproducing the method for data extraction from previous works, we find our Stack Overflow dataset to be 1.5 times larger than the dataset obtained by \cite{yang2016}, and only 0.4\% of our sampled repositories overlap with \cite{zahedi2018}. Additionally, our analysis of the challenges is more thorough as we also consider the popularity, difficulty and expertise. Finally, we perform extensive comparative analysis of the topics across programming languages and data sources. 

\section{Methodology}
\subsection{Research Questions}
This study is designed to answer six major Research Questions (RQs) in order to provide a thorough analysis and understanding of security challenges for different programming languages. 

\begin{itemize}
    \item \textbf{Programming Language Security Consideration.} We first aim to provide a general indication of how comparatively important security is for different languages. We examine the overall amount of discussion for all languages to help select the most popular programming languages for analysis in subsequent research questions. 

    \begin{itemize}
        \item [\textbf{RQ1.}] \textbf{What is the rate of security discussion amongst programming languages on Stack Overflow and GitHub?}\\
        This first research question aims to give a broad view of the security consideration for a large number of languages, via the rate at which security-related discussions occur.  
    \end{itemize}

    \item \textbf{Programming Language Security Discussions.} We answer three research questions related to the content of security discussions. Specifically, we aim to identify what security challenges are faced by developers for different programming languages and why. 
    
    \begin{itemize}
        \item [\textbf{RQ2.}] \textbf{What is the intention behind security discussions for different programming languages?} \\
        This RQ seeks to identify why developers are discussing security topics for particular languages. We analyse the intent, motivations and purpose of the security questions and issues being posted on Stack Overflow and GitHub. Through this knowledge we can better determine why security is discussed differently, and the potential security information needs of developers. 

        \item [\textbf{RQ3.}] \textbf{What are the major security challenges and topics discussed by developers for different programming languages?}\\
        Through RQ3 we aim to identify and compare the major categories of security challenges that developers face when using different programming languages. We also analyse the relative significance of these categories through their prevalence in developer discussions. Through this synthesis of publicly available knowledge, we enable practitioners to better understand the security issues commonly encountered for different languages, and provides researchers with a taxonomy of key security topics. 

        \item [\textbf{RQ4.}] \textbf{How do security discussion topics change over time for different programming languages?}\\
        RQ4 aims to further this analysis by examining how the topics evolve over time, which provides insights into whether certain challenges become more or less important for specific languages as they update. RQ4 enables both practitioners and researchers to observe the relative trends and importance of the different security topics for different languages. 
    \end{itemize}

    \item \textbf{Programming Language Security Support.} Finally, we answer two research questions relating to the community support available for security challenges of different programming languages. We aim to identify the characteristics of how security related challenges are received and handled by the community. To achieve this, we use quantitative metrics derived from discussion metadata for popularity, difficulty and expertise, inspired from previous works \citep{yang2016,ahmed2018,tian2019}. 

    \begin{itemize}
        \item [\textbf{RQ5.}] \textbf{What are the characteristics in terms of popularity and difficulty of different programming languages and their identified security challenges?}\\
        RQ5 uses document-centric metrics to identify the relative popularity (engagement and interest) and difficulty (complexity and  labour). The combination of these metrics gives an indication of how significant the security challenges are when they occur. 

        \item [\textbf{RQ6.}] \textbf{What is the level of security expertise of the users who answer security related discussions for different programming languages?}\\
        RQ6 utilizes user-centric metrics to examine the relative expertise (knowledge and skill) of users of different programming languages. These characteristics help garner further insights into how the challenges are handled, i.e., is there sufficient knowledge and enough experts to help resolve security challenges. As such, we only consider the users who provide answers to the security related discussions. 
    \end{itemize}

\end{itemize}


\subsection{Data Collection}
\begin{figure*}[h]
    \includegraphics[width=\textwidth]{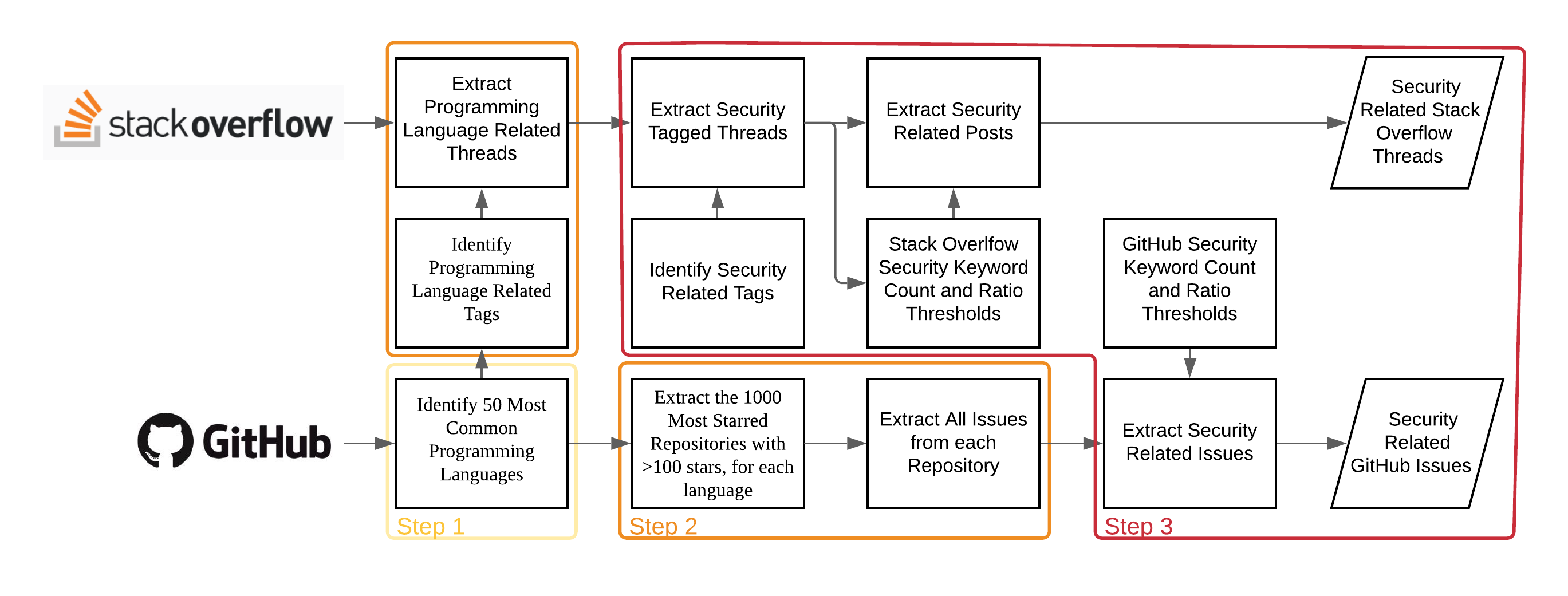}
    \caption{Workflow of retrieving security related posts and issues from Stack Overflow and GitHub}
    \label{fig:data_workflow}
\end{figure*}

To conduct our study we collected a dataset of security and programming language related discussions from Stack Overflow and GitHub.  The workflow for the data extraction process is shown in Fig. \ref{fig:data_workflow}. Our data collection process follows three major steps. First, we identified the programming languages commonly used by developers so that we could scope our study. Secondly, we extracted a dataset of discussions relating to each programming language from both GitHub and Stack Overflow. Thirdly, we filtered the datasets from step 2 to security-related discussions. Step 2 and 3 are performed separately so that we can identify the rate of security discussion as well as more general programming language discussion characteristics. 

To first identify the programming languages used by developers, we selected the 50 most commonly used programming languages on GitHub in 2020 for consideration, as reported by \emph{GitHut}\footnote{\url{https://madnight.github.io/githut}}. We then obtained a programming language specific dataset for each of these 50 programming languages from Stack Overflow and GitHub. 

For Stack Overflow, we extracted programming language specific data based on the tags of each post, similar to prior works \citep{yang2016}. For the 50 identified programming languages, we manually identified the appropriate Stack Overflow tag, as well as all tags relating to different versions and releases of that language. The complete tag list can be viewed in our online appendix\footnote{\url{https://github.com/RolandCroft/Language-Security-Challenges/tree/master/Appendix/SOLangsTop50.txt}}. Every post on Stack Overflow that contains one or more of these tags is extracted; considering all data up to May 1st 2020. 

If a post contains more than one language tag, it is included in each of the relevant language datasets. However, C and C++ are both syntactically and semantically similar, and have heavy overlap in their secure coding properties \citep{seacord2005}. Hence, this approach causes redundancy in the individual datasets of C or C++. The C dataset contain 2641 posts in comparison to the C++ dataset that contains 3136 posts. However, 323 of the 2641 C posts (12\%) are also contained in the C++ Stack Overflow dataset. Furthermore, repositories primarily written in C often also contain C++ code.  Hence, to reduce the redundancy in our analysis, we consider the dataset collected for C to sufficiently represent both the C and C++ language, due to their natural overlap and similar semantic properties. This redundancy was later confirmed in our results. When extracting topics for the C and C++ datasets individually, the six topic categories that manifest for C Stack Overflow posts are also the 6 most dominant topic categories for C++ Stack Overflow posts.

For GitHub, we extracted language specific data based on the predominant language of the repository. However, due to the substantially larger size of GitHub data, we were unable to efficiently analyze all repositories. Hence, we decided to use a repository sampling method similar to the previous work by \cite{zahedi2018}. We retrieved the top 1000 most starred repositories for each programming language. To ensure that we only sampled high quality repositories for large scale software projects, only repositories with more than 100 stars are kept from the top 1000. For GitHub we extended the data collection time span slightly to increase the amount of available data for each repository, considering data up to July 1st 2020. 

Thirdly, we filtered the programming language datasets to security-related discussions. As we aim to identify the security challenges that are discussed, it is important that the discussions in our dataset are relevant and focused towards security. To achieve this, we utilized both tag-based filtering and content-based filtering on our datasets; where applicable. 

\begin{figure}[h]
\centering
    \includegraphics[width=0.4\textheight]{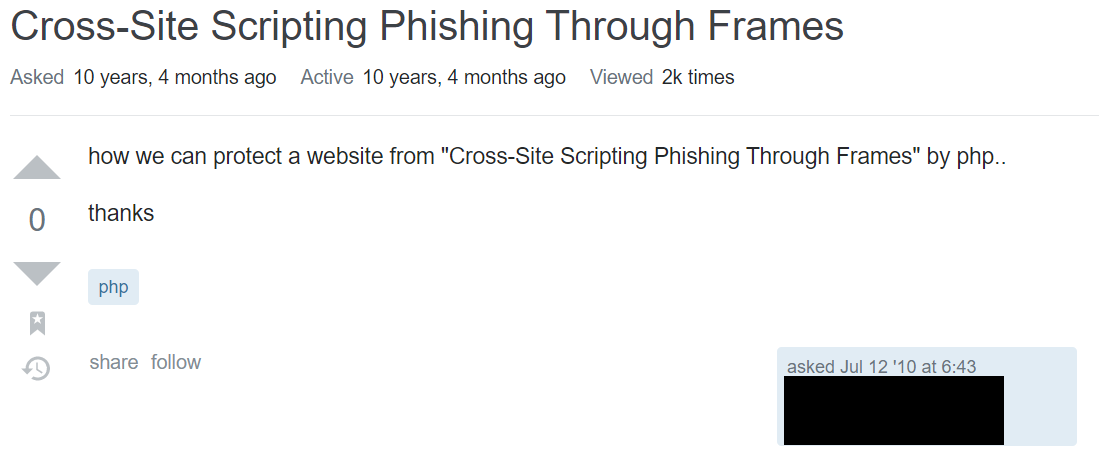}
    \caption{A Stack Overflow post (ID: 3226374) with incomplete tags}
    \label{fig:untagged_so}
\end{figure}

Tag-based filtering uses the tags of a post to identify the category of its content. This approach is most commonly adopted by previous works \citep{yang2016, ahmed2018, bagherzadeh2019} due to its simplicity and precision, but it still requires the manual identification of appropriate tags. However, we cannot rely upon this approach alone as a reliable tagging system does not exist for GitHub issues. Furthermore, tagging itself is unreliable as it is performed by the user, who can submit incorrect or insufficient labels. For example, Fig. \ref{fig:untagged_so} depicts a Stack Overflow post which is not tagged with any security tag despite stimulating discussion about cross-site scripting and phishing. 

Content-based filtering examines the actual content of a post to determine its relevance to a particular category through a set of keywords. If a post contains a lot of keywords which are representative of a particular topic (i.e., security), then the entire post can be considered as related to this topic. Content-based filtering has also been used in previous works when tag-based filtering could not be applied \citep{pletea2014, zahedi2018} or to account for the shortcomings of tag-based filtering \citep{le2020}. However, as keyword matching can produce false positives, content-based filtering requires threshold values to determine a post as relevant or not. For example, we cannot consider any post containing the word \emph{cookie} as relating to web development, as this word may instead be referring to the tasty baked treat. Thus, we use unique count (the number of unique keywords in a post) and ratio (the ratio of keywords to the total number of words in a post) thresholds to ensure discussion relevance. 

We use both the tag-list and keyword-list proposed by \cite{le2020}, as they are some of the most extensive security keyword lists from the previous literature. Our tag list consists of the tags: \emph{security, cryptography, csrf, passwords, sql-injection} and
\emph{xss}, with the \emph{security} tag being sub-string matched and all others exact matched. Our  keyword list originally contained 234 security keywords, which we updated with manually identified missing variations to a total of 288 keywords\footnote{\url{https://github.com/RolandCroft/Language-Security-Challenges/tree/master/Appendix/security\_keywords.txt}}. 

For content-based filtering all keywords and text are stemmed before keyword matching is applied. We consider different matching forms (i.e., American/British and with/without hyphen/space/stemming) to cope with various spellings.  To reduce false positives we performed exact matching for short (three-character) keywords (e.g., md5) and sub-word matching otherwise \citep{le2020}. 

We empirically identified the suitable content thresholds for content-based filtering. For Stack Overflow, we obtained the content thresholds from the median values of the security tag filtered dataset; unique count = 4, ratio = 0.115. For GitHub, we first obtained all posts that contain one or more of the security keywords in the title, similar to the method used by \cite{zahedi2018} and \cite{pletea2014}, and then manually examine a sample of 385 of these issues. Of this sample, only 70\% of the posts were able to be confidently labelled as security-related by the authors, highlighting the need for stricter content thresholds. A range of threshold values were tested until a sample was deemed as sufficiently security related ($>$95\% security-related); which uses a threshold of unique count = 2, ratio = 0.061. 

Finally, we form our dataset of security related discussions for each language from each data source. For Stack Overflow, we used the union (non-overlapping) of the tag-based filtering set and content-based filtering set. For GitHub, we only used the content-based filtering set as previously stated. 

We manually examined a sample of documents to confirm the content and security relation of the datasets\footnote{\url{https://github.com/RolandCroft/Language-Security-Challenges/tree/master/Appendix/dataset\_samples}}. We selected 385 documents from each source, for a confidence level of 95\% and a confidence interval of 5\% \citep{cochran2007}. For Stack Overflow 100\% of the examined posts have a relation to security. For GitHub, 95\% of the issues are considered as security relevant, while the other 5\% contain unclear content or no reference to security. 

For RQ1 we examine the collected data for all 50 languages. However, for subsequent research questions we limit our study to the most popular programming languages which have a substantial security consideration, to ensure that there is a sufficient amount of data to analyse and provide a better comparison. We select these languages as any language with a statistically significant number (\textgreater384) \citep{cochran2007} of security related discussion posts on both Stack Overflow and GitHub. Hence, we only perform our in-depth analysis of the security challenges on a refined subset of 15 of the most dominant programming languages, shown in Table \ref{table:num_topics}. 

\subsection{Manual Discussion Analysis}
To address RQ2, we perform manual analysis of our data to determine the nature and purpose of the discussions. As the entirety of our dataset is naturally too large to manually examine, we take a statistically significant sample (confidence level 90\% +/-10\%) \citep{cochran2007} of 68 posts per language for each data source (n=1904). We then manually categorize the purpose of each discussion based on the content of the question or issue. To define and categorize discussion intent, we adopt the curated taxonomy of question categories curated by \cite{beyer2020}. This taxonomy consists of seven question categories, as defined by \cite{beyer2020} in Table \ref{table:question_cats}. 

\begin{table}[h]
\centering
\caption{The seven question categories as defined by \cite{beyer2020} }
\label{table:question_cats}
\resizebox{\columnwidth}{!}{%
\begin{tabular}{ |p{2cm}|p{10cm}| } 
    \hline
    \textbf{Category} & \textbf{Definition}\\ 
    \hline
    API Usage & ``The posts falling into this category contain questions asking for suggestions on how to implement some functionality or how to use an API. The questioner is asking for concrete instructions.''\\ 
    \hline
    Discrepancy & ``The posts of this category contain questions about problems and unexpected behavior of code snippets whereas the questioner has no clue how to solve it.''\\ 
    \hline
    Errors & ``Similar to the previous category, posts of this category deal with problems of exceptions and errors. Often, the questioner posts an exception and the stack trace and asks for help in fixing an error or understanding what the exception means.''\\
    \hline
    Review & ``Questioners of these posts ask for better solutions or reviewing of their code snippets. Often, they also ask for best practice approaches or ask for help to make decisions, for instance, which API to select.''\\
    \hline
    Conceptual & ``The posts of this category consist of questions about the limitations of an API and API behavior, as well as about understanding concepts, such as design patterns or architectural styles, and background information about some API functionality.''\\
    \hline
    API Change & ``These posts contain questions that arise due to the changes in an API or due to compatibility issues between different versions of an API.''\\
    \hline
    Learning & ``In these posts, the questioners ask for documentation or tutorials to learn a tool or language. In contrast to the first category, they do not aim at asking for a solution or instructions on how to do something. Instead, they aim at asking for support to learn on their own.''\\
    \hline
\end{tabular}%
}
\end{table}

As GitHub issues discussions do not strictly follow the question/answer format of Stack Overflow, we extend the taxonomy slightly to additionally account for the intent of GitHub user discussions. Firstly, GitHub issues related to Bug Fixes or inclusion of missing critical functionality are classified with the \textit{Errors} category, as these issues are discussing fixes for software errors. Secondly, GitHub issues which introduce new or additional functionality (i.e., updates), are classified as \textit{Review} as these issues are requesting for review, feedback or approval of code. Otherwise, we follow the same manual classification technique as described by \cite{beyer2020}. A post may relate to more than one category depending on its content. 

Two of the authors manually categorized each post to the above question categories. The authors first independently labelled 100 posts (50 from each source) to ensure consistent labelling. We calculated the Cohen's kappa coefficient \cite{cohen1960} to determine the inter-rater agreement of the two authors, by calculating the kappa score for each label separately and then averaging scores. An overall kappa score of 0.65 was achieved on the initial set, implying substantial inter-rater agreement. Any initial disagreements in labelling were resolved through discussion. The authors then individually labelled the remaining posts. 

\subsection{Topic Modelling}
We used topic modelling to identify the security discussion topics in our data sources due to its wide use on similar datasets and in related studies for both Stack Overflow \citep{yang2016, barua2014, zou2015, allamanis2013} and GitHub \citep{zahedi2018, rahman2014}. LDA exhibits the ability to form models without the need for labelled training data or pre-defined taxonomies. This suits our needs as Stack Overflow and GitHub discussions are largely unlabelled, and these informal discussions do not align with formal security taxonomies. 

\begin{figure*}[h]
    \includegraphics[width=\textwidth]{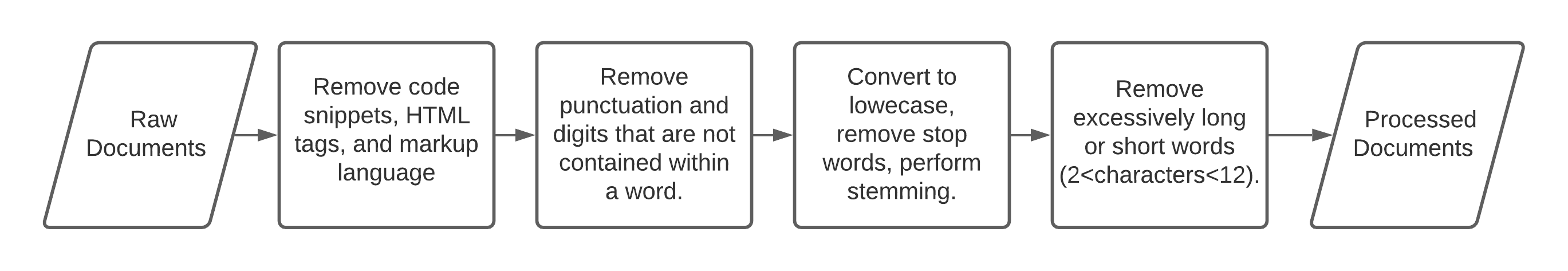}
    \caption{Pre-processing steps of Stack Overflow and GitHub documents}
    \label{fig:preprocessing}
\end{figure*}

We used the Python \emph{Gensim} library for the implementation of the LDA algorithm\footnote{\url{https://radimrehurek.com/gensim/}}. As Stack Overflow and GitHub discussions are naturally noisy \citep{barua2014, zahedi2018}, the pre-processing steps in Fig. \ref{fig:preprocessing} are applied. We use the union of the \emph{Sklearn} and \emph{NLTK} stopword lists \citep{nothman2018}, and the \emph{NLTK} Porter-Stemmer Algorithm. 

\cite{treude2019} show that corpora sampled from different languages of GitHub and Stack Overflow have varying characteristics and require different parameter configurations. Hence, we consider each language-specific security related dataset from each data source as a separate corpora, and trained the LDA models individually. In a related study, \cite{han2020} also adopt a similar approach that they title as Balanced LDA. This method also removes any potential influence from the varying number of documents in each corpora. 

Topic model quality heavily relies on appropriate hyper-parameter tuning \citep{agrawal2018}. Hence, we optimized the following three LDA hyper-parameters via grid search: the number of topics \emph{k}, the a-priori belief of each topic probability $\alpha$, and the a-priori belief of each word probability $\beta$. For \emph{k}, all values in the range 1 to 30 are tested. For $\alpha$ and $\beta$, we tested the default configuration ($\alpha=\beta=1/k$), the automatically optimized configuration by \emph{Gensim}, and the standard values from previous literature of $\alpha=50/k$ and $\beta=0.01$ \citep{ahmed2018, bagherzadeh2019, yang2016}. A model is trained with every combination of settings for each corpora. 

Topic coherence measures the semantic similarity of high probability words in a topic \citep{mimno2011}. This metric is selected for model evaluation, due to its ability to predict topic interpretability \citep{mimno2011} and its prevalence in related studies \citep{ahmed2018, wan2019}. The optimal model is chosen as the model with the highest coherence value. In the event of similar coherence values, the lower number of topics is chosen as it was empirically found that higher topic numbers produce more fine-grained and less comparable topics. The optimal number of topics for each corpora is shown in Table \ref{table:num_topics}. 

\begin{table}[h]
\centering
\caption{The number of security-related Stack Overflow posts or GitHub issues and the optimal number of topics for each corpora}
\label{table:num_topics}
\resizebox{\columnwidth}{!}{%
\begin{tabular}{ |c|c|c|c|c|c|c| } 
    \hline
    \makecell{Programming\\Language} & \makecell{\# Stack Over-\\flow Posts} & \makecell{\# Stack Over-\\flow Topics} & \makecell{Stack Over-\\flow Model\\Coherence} & \makecell{\# GitHub\\Issues} & \makecell{\# GitHub\\Topics} & \makecell{GitHub Model\\Coherence}\\ 
    \hline
    Java & 35576 & 17 & 0.516 & 13647 & 14 & 0.538\\ 
    PHP & 32191 & 17 & 0.498 & 20398 & 19 & 0.502\\ 
    JavaScript & 22624 & 22 & 0.461 & 16246 & 17 & 0.489\\
    C\# & 22151 & 23 & 0.485 & 11892 & 26 & 0.495\\
    Python & 11353 & 18 & 0.515 & 15368 & 20 & 0.566\\
    Shell & 5807 & 23 & 0.443& 6395 & 11 & 0.567\\
    C/C++ & 2641 & 13 & 0.472 & 16794 & 19 & 0.545\\
    Ruby & 2514 & 9 & 0.450 & 14537 & 11 & 0.535\\
    Objective-C & 2093 & 3 & 0.439 & 2472 & 14 & 0.446\\
    Swift & 1859 & 21 & 0.428 & 2124 & 7 & 0.522\\
    PowerShell & 1782 & 7 & 0.384 & 4112 & 3 & 0.584\\
    Perl & 942 & 11 & 0.442 & 2188 & 2 & 0.582\\ 
    TypeScript & 856 & 8 & 0.437 & 10227 & 24 & 0.526\\ 
    Go & 809 & 3 & 0.402 & 24288 & 13 & 0.507\\ 
    \hline
\end{tabular}%
}
\end{table}

From the output of LDA, we obtain a set of $K$ topics ($z_1,...,z_k$) from each of our $N$ corpora ($c_1,...,c_n$). The symbols used to represent the topic attributes are summarized in Table \ref{table:topic_legend}. Each topic is a distribution over the unique keywords of each corpora. For each post $d$ in the corpora we also obtain a topic probability vector which indicates the proportion of words that come from each of the $K$ topics \citep{blei2003}. We denote the probability of a particular topic $z_k$ for a particular post $d_i$ as $\theta(d_i, z_k)$. Note that $\forall i,k : 0 \leq \theta(d_i, z_k) \leq 1$ and $\forall i : \sum_1^k \theta(d_i, z_k) = 1$. 

\begin{table}[h]
\centering
\caption{Topic attribute symbols}
\label{table:topic_legend}
\resizebox{\columnwidth}{!}{%
\begin{tabular}{ |p{1.5cm}|p{4cm}|p{6cm}| } 
    \hline
    \textbf{Symbol} & \textbf{Meaning} & \textbf{Specifics}\\ 
    \hline
    ($z_k$) & Topic \emph{k} & A security topic. \\
    \hline
    ($c_j$) & Corpora \emph{j} & A security related programming language dataset from Stack Overflow or GitHub. \\
    \hline
    ($d_i$) & Document \emph{i} & A Stack Overflow post or GitHub issue. \\
    \hline
    $\theta(d_i, z_k)$ & Document-Topic Probability & The probability of a document belonging to a topic. \\ 
    \hline
\end{tabular}%
}
\end{table}

Using the topic word distribution and topic probability vectors, topic labels are then assigned through collaboration of two of the authors. The authors assign topic categories using an open card sorting method \citep{fincher2005}, in which the authors generate a dendogram of the items using self-generated labels. The authors first use the topic words to infer the label based on their own security expertise and judgement. The authors then examine documents in order of $\theta(d_i, z_k)$ (document-topic probability) to validate the inferred label. The authors examine documents until they are confident with the label; at least 10 documents were examined for each topic (provided the topic has more than 10 associated documents). If the authors are unconfident with the label after an examined portion of posts, a larger portion is examined to help generate a new label. Topics that do not implicitly relate to security or are too ambiguous to label are labelled as \textit{Other}. 

The first two authors first independently examined all topics to familiarize themselves with the data and gain an understanding of the potential topics and themes. During this process, each author tentatively assigned labels to each topic. The first two authors then collaboratively discussed and refined the topic categories. Finally, to reduce potential individual researcher bias during this manual process, the final topic labels are assigned collaboratively by the first two authors in unison. The first author has over three years of experience doing security related research in academia, whereas the second author has over 13 years of security experience working in industry. Disagreements are resolved through discussion. 

To assess the potential subjectivity of the topic labelling process, we used an approach similarly performed in other works \citep{hata20199}. Once, the final topic categories had been defined, the first two authors independently annotated a random sample of 30 topics. We then calculated the Cohen's kappa coefficient \citep{cohen1960} to determine the inter-rater agreement of the two authors, by calculating the kappa score for each topic category separately and then averaging scores. An overall kappa score of 0.79 was achieved on the initial set, implying substantial inter-rater agreement. Based on this encouraging result, we considered the topic labelling process to be sufficiently robust to potential researcher bias.

After all labels have been generated, topic grouping and categorisation is then conducted using a bottom-up categorisation approach: beginning with the initially assigned specific topic labels, sub-categories of topics are then identified, which are then further grouped into final high level categories. Whilst individual topics and sub-categories can potentially relate to numerous groupings, we assign them to the most relevant category. To reduce the dimension of the data, we only consider (sub-)categories for analysis, and group individual topics based on this. 

We provide a full example of the labelling process here. One of the produced topics for Java contains the keywords \textit{user, authentication, application, ldap, credentials, access, login}. Based on the context of these keywords we interpret the topic as being related to user login and authentication. We confirm this interpretation by then examining associated posts for this topic to see if they fit the context, e.g., \textit{``How can I allow my users to login with Windows credentials?'' (SO, 13927479)}. Once we have completed this process, we then consider all topics, regardless of source and programming language, to create our topic taxonomy. For each topic, we first assign a category based on other similar topics that have manifested. Topics that have many other semantically similar topics become a category, and are assigned a label that broadly captures the definition of all the topics. This example topic we assign the Authentication category. Then once again after we have assigned all topics to a category, we finally group the categories into higher-level definitions; in this case, Access Control. 

The security taxonomy that emerged from our data is presented in Table \ref{table:topic_numbers}. We also show the number of individual topics that was mapped to each category. 

\begin{table}[h]
\centering
\caption{Our identified security taxonomy and topic mapping }
\label{table:topic_numbers}
\begin{tabular}{ |c|c|c| } 
    \hline
    \textbf{Category} & \textbf{Sub-Category} & \textbf{\# Topics}\\ 
    \hline
    \multirow{2}{*}{Cryptography} & Encryption & 42\\
    \cline{2-3}
     & Encoding & 12\\
    \hline
    \multirow{3}{*}{Access Control} & Authentication & 40\\
    \cline{2-3}
     & Tokens & 25\\
    \cline{2-3}
     & Authorization & 10\\
     \hline
    \multirow{3}{*}{Network Security} & Digital Certificates & 26\\
    \cline{2-3}
     & Client/Server & 23\\
    \cline{2-3}
     & File Transfer & 7\\
    \hline
    \multirow{2}{*}{Data Security} & Password/Data Storage & 23\\
    \cline{2-3}
     & Digital Signature & 7\\
    \hline
    \multirow{4}{*}{Software Vulnerabilities} & Network Exploits & 18\\
    \cline{2-3}
     & Mitigation & 5\\
    \cline{2-3}
     & Memory Management & 6\\
    \cline{2-3}
     & Race Condition & 4\\
    \hline
    \multirow{4}{*}{Secure Development} & Implementation & 57\\
    \cline{2-3}
     & Testing & 25\\
    \cline{2-3}
     & Libraries/Configuration & 39\\
    \cline{2-3}
     & GitHub/Reports & 15\\
    \hline
    \multicolumn{2}{|c|}{Other} & 11\\
    \hline
     \multicolumn{2}{|c|}{\textbf{Total}} & \textbf{395}\\
    \hline
\end{tabular}
\end{table}

We define the dominance of a topic in a particular corpus via the \emph{Topic Share} metric \citep{barua2014}. The topic share value measures the proportion of posts in a corpus ($c_j$) which contain a specific topic ($z_k$):
\begin{equation}
    share(z_k, c_j) = 100 \times \frac{1}{|c_j|} \sum_{d_i\in c_j} \theta(d_i, z_k)
\end{equation}

To evaluate the temporal trends of particular topics within the corpora for RQ4, we utilize the topic impact metric, as used in similar works \citep{panichella2013, rosen2016, wan2019}. The impact of a topic $z_k$ in month $m$ is defined as:
\begin{equation}
    impact (z_k, m) = \frac{1}{|D(m)|} \sum_{d_i\in D(m)} \theta(d_i, z_k)
\end{equation}
where $D(m)$ is the set of documents within the corpus that were posted in month $m$. For example, for a month in which 3 issues were posted which each  relate to the same topic with probabilities: $0.3, 0.4, 0.8$, then the topic impact value of that topic for that month would be: $(0.3+0.4+0.8)/3 = 0.5$. This gives us an indication of the each topic's significance per month, for the time range of January 2009 to May 2020. The topic impact score can be considered as the topic share value over time.  

\subsection{Topic and Language Characteristics}

We obtain various metrics of the posts and users for each language and its topic categories to reveal aspects of their characteristics. These metrics inform us of the relative significance of the challenges when they occur, and the nature in which they are resolved by online communities. 

For RQ5, we focus on the attributes of the documents that make up the various topics and languages. As it is difficult to determine the document-centric metrics for partial topics via the topic share value, we cluster posts based on their topic probability vectors. We define the dominant topic for each post to be the topic with the highest probability for that document \citep{panichella2013, rosen2016, wan2019}. This is formally defined as:
\begin{equation}
    dominant(d_i) = z_k : \theta(d_i, z_k) = max(\theta(d_i, z_j)), 1 \leq j \leq K
\end{equation}
The size of a topic ($|z_k|$) is the number of posts for which it is the dominant topic, i.e., $\sum max(\theta(d_i, z_k))$. To obtain the set of dominant topics for a category, we take the union of document sets for each topic which forms that category. Similarly, we can calculate the overall popularity or difficulty of a language through the set of all documents pertaining to that language. 

For Stack Overflow, we use document-centric metrics inspired by previous works \citep{yang2016, bagherzadeh2019, ahmed2018} to measure the popularity and difficulty. 

The popularity of a given Stack Overflow topic category is calculated as follows:
\begin{equation}
    Popularity_{SO} = \frac{1}{\left|T\right|} \sqrt[\leftroot{-2}\uproot{2}4]{\sum P1 \times \sum P2 \times \sum P3 \times \sum P4}
\end{equation}
where for each document in that topic, P1 is the number of views, P2 is the score, P3 is the favorites, P4 is the number of comments, and $\left|T\right|$ is the number of documents in that topic category. 

The difficulty of a given Stack Overflow topic category is calculated as follows:
\begin{equation}
    Difficulty_{SO} = \sqrt[\leftroot{-2}\uproot{2}3]{\frac{\left|T\right|}{\sum D1} \times \frac{\sum D2}{\left|T\right|_{accepted}} \times \frac{\left|T\right|_{accepted}}{\sum D3}}
\end{equation}
where for each document in that topic, D1 is whether the question has an accepted answer, D2 is the time in hours to receive an accepted answer and D3 is the number of answers per the number of views. As not all questions have answers or accepted answers, D2 and D3 are determined in relation only to posts that have an accepted answer. D1 and D3 have an inverse relationship to difficulty; a higher value indicates an easier topic.  

To the best of our knowledge, ours is the first work to attempt to formally quantify the popularity or difficulty of GitHub Issues via their discussions. We adapt the metrics used for Stack Overflow to also be suitable for GitHub documents as well, but due to the differences in the sites metadata and activities we note that the conversion is not directly equivalent. 

We define the popularity of a given GitHub topic category as follows: 
\begin{equation}
    Popularity_{GH} = \frac{1}{\left|T\right|} \sqrt[\leftroot{-2}\uproot{2}3]{\sum P1 \times \sum P2 \times \sum P3}
\end{equation}
where for each document in that topic, P1 is the number of comments, P2 is the number of unique users who contribute to the issue and P3 is the number of 'up' reactions\footnote{\url{https://developer.github.com/v3/reactions/}}. We use GitHub reactions as a metric due to their possible resemblance to the score of a Stack Overflow posts, and reactions are becoming increasingly commonplace in GitHub issue discussions \citep{borges2019}. 

The difficulty of a GitHub topic category is defined as:
\begin{equation}
    Difficulty_{GH} = \frac{1}{\left|T\right|} \sqrt[\leftroot{-2}\uproot{2}2]{\sum D1 \times \sum D2}
\end{equation}
where for each document in that topic, D1 is the time in hours to close that issue and D2 is the number of commits assigned to that issue. We find that there are a significant number of outliers for issue close time due to stale issues \citep{wessel2019}, so for D1 we remove outliers using an outlier boundary of $1.5 * Interquartile Range$. 

It is important to note that these metrics are different to the topic share metric measured in RQ3 and obtain a different ranking, which we confirm using the Kendall rank correlation coefficient \citep{knight1966} at the 95\% confidence level. Whilst the topic share indicates the overall prevalence of a topic, the popularity measures the interest and engagement of a post for that topic when it occurs. 

For expertise, we adopt the metrics proposed by \cite{tian2019} which identify the relative expertise of a user on either Stack Overflow or GitHub for a particular topic. The full equations for the expertise of a Stack Overflow or GitHub user can be seen in their paper \citep{tian2019}. In our case the topic of expertise is \textit{security}. We adopt this particular approach as it considers various important user attributes and platform specific information. 

For Stack Overflow, the expertise of a given user is calculated through a combination of their profile performance (reputation and profile views) and their answer performance (answer scores and score/favorites/views of questions they are answering). To relate the expertise to a specific topic, only the answer performance for questions of that topic are considered. Hence, we obtain the user answer performance for the security related posts. To extend the generalizability of security knowledge, we also consider general security related posts which are collected using the same heuristics as the dataset created by \cite{yang2016}. 

For GitHub, the expertise of a user is determined through their contribution to significant repositories. This is calculated by the frequency of a user's commits, and the total commits and watchers of a repository. To make the expertise value topic specific, the topic weight of the repository is also determined. We altered the weight formula proposed by \cite{tian2019} slightly, to better represent a user's security contribution, and calculate it as:
\begin{equation}
    Weight = \frac{numUserSecurityComments}{numSecurityComments}
\end{equation}

As we intend to investigate how challenges are resolved, only the expertise of users who provide answers to security related discussions are considered, either through the accepted answer on Stack Overflow, or a comment on a GitHub issue. 

The individual values of these metrics (popularity, difficulty and expertise) are not indicative on their own, so we only consider the normalized values for the purposes of comparison in RQ5 and RQ6. 

\section{Results and Analysis}

\subsection{RQ1: What is the rate of security discussion amongst programming languages on Stack Overflow and GitHub?}
\begingroup
\def\arraystretch{0.66}
\begin{table*}[t]
\caption{The rate of security related discussion (\%) for each programming language for each source}
\label{table:discussion_rate}
\resizebox{\textwidth}{0.32\textheight}{%
\begin{tabular}{ |c|c|c|c|c|c|c|c|c| } 
    \hline
    \multirow{2}{*}{\makecell{Programming\\Language}} & \multicolumn{3}{|c|}{Stack Overflow} & \multicolumn{3}{|c|}{GitHub} & \multirow{2}{*}{Difference*} & \multirow{2}{*}{Average}\\
    \cline{2-7}
    & \makecell{Security\\Posts} & \makecell{Total\\Posts} & \makecell{Security\\Rate} & \makecell{Security\\Issues} & \makecell{Total\\Issues} & \makecell{Security\\Rate} & & \\
    \hline
	Shell & 5807 & 152203 & \textbf{3.82} & 6395 & 458529 & \textbf{1.39} & +2.42 & \textbf{2.60}\\
	\hline
	PHP & 32191 & 1361990 & \textbf{2.36} & 20398 & 1157572 & \textbf{1.76} & +0.60 & \textbf{2.06}\\
	\hline
	PowerShell & 1782 & 82460 & \textbf{2.16} & 4112 & 266222 & \textbf{1.54} & +0.62 & \textbf{1.85}\\
	\hline
	Go & 809 & 46051 & \textbf{1.76} & 24288 & 1448882 & \textbf{1.68} & +0.08 & \textbf{1.72}\\
	\hline
	Erlang & 126 & 8816 & \textbf{1.43} & 1495 & 78827 & \textbf{1.90} & -0.47 & \textbf{1.66}\\
	\hline
	Java & 35576 & 1689665 & \textbf{2.11} & 13647 & 1234379 & \textbf{1.11} & +1.00 & \textbf{1.61}\\
	\hline
	Puppet & 49 & 3762 & \textbf{1.30} & 198 & 11734 & \textbf{1.69} & -0.38 & \textbf{1.49}\\
	\hline
	Perl & 942 & 65995 & \textbf{1.43} & 2188 & 145351 & \textbf{1.51} & -0.08 & \textbf{1.47}\\
	\hline
	C/C++ & 2641 & 330139 & \textbf{0.80} & 16794 & 871251 & \textbf{1.93} & -1.13 & \textbf{1.36}\\
	\hline
	Ruby & 2514 & 214226 & \textbf{1.17} & 14537 & 987441 & \textbf{1.47} & -0.30 & \textbf{1.32}\\
	\hline
	C\# & 22151 & 1432299 & \textbf{1.55} & 11892 & 1118533 & \textbf{1.06} & +0.48 & \textbf{1.30}\\
	\hline
	Groovy & 329 & 25317 & \textbf{1.30} & 1259 & 119471 & \textbf{1.05} & +0.25 & \textbf{1.18}\\
	\hline
	Python & 11353 & 1536950 & \textbf{0.74} & 15368 & 963333 & \textbf{1.60} & -0.86 & \textbf{1.17}\\
	\hline
	VB.NET & 1332 & 131530 & \textbf{1.01} & 55 & 4655 & \textbf{1.18} & -0.17 & \textbf{1.10}\\
	\hline
	Elixir & 82 & 7737 & \textbf{1.06} & 1145 & 108948 & \textbf{1.05} & +0.01 & \textbf{1.06}\\
	\hline
	JavaScript & 22624 & 2016932 & \textbf{1.12} & 16246 & 2080976 & \textbf{0.78} & +0.34 & \textbf{0.95}\\
	\hline
	Objective-C & 2093 & 291030 & \textbf{0.72} & 2472 & 270107 & \textbf{0.92} & -0.20 & \textbf{0.82}\\
	\hline
	Kotlin & 298 & 39886 & \textbf{0.75} & 2011 & 236977 & \textbf{0.85} & -0.10 & \textbf{0.80}\\
	\hline
	Lua & 111 & 17017 & \textbf{0.65} & 1340 & 143203 & \textbf{0.94} & -0.28 & \textbf{0.79}\\
	\hline
	CoffeeScript & 68 & 9767 & \textbf{0.70} & 933 & 105551 & \textbf{0.88} & -0.19 & \textbf{0.79}\\
	\hline
	Swift & 1859 & 285658 & \textbf{0.65} & 2124 & 248266 & \textbf{0.86} & -0.20 & \textbf{0.75}\\
	\hline
	Clojure & 82 & 16297 & \textbf{0.50} & 1236 & 123395 & \textbf{1.00} & -0.50 & \textbf{0.75}\\
	\hline
	Fstar & 0 & 20 & \textbf{0} & 39 & 2642 & \textbf{1.48} & -1.48 & \textbf{0.74}\\
	\hline
	TypeScript & 856 & 114778 & \textbf{0.75} & 10227 & 1455275 & \textbf{0.70} & +0.04 & \textbf{0.72}\\
	\hline
	Assembly & 111 & 34129 & \textbf{0.33} & 252 & 27049 & \textbf{0.93} & -0.61 & \textbf{0.63}\\
	\hline
	QML & 20 & 9094 & \textbf{0.22} & 85 & 8265 & \textbf{1.03} & -0.81 & \textbf{0.62}\\
	\hline
	Dart & 194 & 29978 & \textbf{0.65} & 1253 & 212221 & \textbf{0.59} & +0.06 & \textbf{0.62}\\
	\hline
	Rust & 44 & 16243 & \textbf{0.27} & 4568 & 498734 & \textbf{0.92} & -0.65 & \textbf{0.59}\\
	\hline
	Pascal & 10 & 2156 & \textbf{0.46} & 187 & 25911 & \textbf{0.72} & -0.26 & \textbf{0.59}\\	
	\hline
	Groff & 0 & 56 & \textbf{0} & 77 & 6783 & \textbf{1.14} & -1.14 & \textbf{0.57}\\
	\hline
	Haskell & 105 & 43795 & \textbf{0.24} & 1378 & 165280 & \textbf{0.83} & -0.59 & \textbf{0.54}\\
	\hline
	Vala & 3 & 876 & \textbf{0.34} & 166 & 23085 & \textbf{0.72} & -0.38 & \textbf{0.53}\\
	\hline
	Scala & 327 & 97971 & \textbf{0.33} & 3203 & 456422 & \textbf{0.70} & -0.37 & \textbf{0.52}\\
	\hline
	TSQL & 221 & 62276 & \textbf{0.35} & 246 & 47713 & \textbf{0.52} & -0.16 & \textbf{0.44}\\
	\hline
	OCaml & 5 & 6206 & \textbf{0.08} & 698 & 98536 & \textbf{0.71} & -0.63 & \textbf{0.39}\\
	\hline
	Smalltalk & 4 & 1563 & \textbf{0.26} & 63 & 12573 & \textbf{0.50} & -0.25 & \textbf{0.38}\\
    \hline
	VimL & 0 & 41 & \textbf{0} & 255 & 121604 & \textbf{0.72} & -0.72 & \textbf{0.36}\\
	\hline
	D & 3 & 2540 & \textbf{0.12} & 259 & 47265 & \textbf{0.55} & -0.43 & \textbf{0.33}\\
	\hline
	DM & 0 & 49 & \textbf{0} & 736 & 121218 & \textbf{0.61} & -0.61 & \textbf{0.30}\\
	\hline
	F\# & 17 & 14631 & \textbf{0.12} & 211 & 46838 & \textbf{0.45} & -0.33 & \textbf{0.28}\\
	\hline
	Scheme & 2 & 7196 & \textbf{0.03} & 57 & 10966 & \textbf{0.52} & -0.49 & \textbf{0.27}\\
	\hline
	Raku & 0 & 395 & \textbf{0} & 41 & 9101 & \textbf{0.45} & -0.45 & \textbf{0.23}\\
	\hline
	Prolog & 9 & 11246 & \textbf{0.08} & 3 & 883 & \textbf{0.34} & -0.26 & \textbf{0.21}\\
	\hline
	ELisp & 3 & 3735 & \textbf{0.08} & 415 & 132570 & \textbf{0.31} & -0.23 & \textbf{0.20}\\
	\hline
	R & 320 & 346073 &\textbf{0.09} & 487 & 166897 & \textbf{0.29} & -0.20 & \textbf{0.19}\\
	\hline
	Coq & 2 & 2053 & \textbf{0.10} & 14 & 7658 & \textbf{0.18} & -0.09 & \textbf{0.14}\\
	\hline
	Julia & 6 & 6757 & \textbf{0.09} & 255 & 149348 & \textbf{0.17} & -0.08 & \textbf{0.13}\\
	\hline
	Fortran & 1 & 10903 & \textbf{0.01} & 51 & 34087 & \textbf{0.15} & -0.14 & \textbf{0.08}\\
	\hline
	MATLAB & 67 & 88285 & \textbf{0.08} & 3 & 18230 & \textbf{0.02} & +0.06 & \textbf{0.05}\\
    \hline
    \hline
    Total & 150285 & 11358546 & \textbf{1.32} & 235141 & 20832355 & \textbf{1.13} & 0.19 & \textbf{1.23}\\
    \hline
\end{tabular}%
}
\tiny \textbf{*} A + symbol indicates the increase in rate of discussion on Stack Overflow compared to GitHub, and a - symbol indicates a decrease in rate of discussion on Stack Overflow compared to GitHub.  
\end{table*}
\endgroup

The selected languages and their rate of security discussion is shown in Table \ref{table:discussion_rate}. Related literature has reported the frequency of security discussion on Stack Overflow to be between 1\% \citep{yang2016} and 2\% \citep{le2020}, and the rate of security discussion on GitHub to be between 3\% \citep{zahedi2018} and 10\% \citep{pletea2014}. For our dataset, we observe the average rate of security discussion on Stack Overflow to be 1.32\% which is inline with previous work. However, for GitHub we observe a lower average rate of 1.13\%. Using the same data filtering methods as \cite{zahedi2018} on our dataset, we identify a rate of 3.38\%, but we occasionally found these posts to lack relevance and focus. Hence, we used stricter keyword count thresholds for content-based filtering to ensure document quality, which results in the lower rate of discussion. 

Shell-based languages (i.e., Shell and PowerShell) have some of the highest rates of security discussion (2.6\% and 1.85\%), which is likely due to their stronger coupling with the underlying operating system. Languages commonly used for web development (i.e., PHP, Ruby and JavaScript) also have slightly higher rates of security discussion (2.06\%, 1.32\% and 0.95\%) due to the increased security needs of these highly connected software systems. Conversely, languages oriented towards numerical analysis (Scientific Programming Languages, i.e., R, Julia and MATLAB) have exceedingly low rates of security discussion with less than 0.2\% each. 

General purpose languages typically have a higher average rate of security discussion, and more obscure languages have a lower average rate of security discussion. Similarly, several of the languages with the highest rates of security discussion are widely used (i.e., Go, Java, C/C++, C\# and Python). However, these factors may only be because of their naturally higher rate of overall discussion and use. 

Interestingly, languages which require manual memory management (i.e., C/C++, Objective-C, Pascal, Assembly and Fortran) do not necessarily have a higher rate of security discussion compared to languages which are considered``memory-safe''. Programs written in these languages are especially prone to extreme and unexpected input data via buffer overflow and other related vulnerabilities. However, C/C++ exhibits the highest frequency of security discussion for GitHub issues (1.93\%). 

Thirty nine out of the fifty examined languages exhibit a cumulative rate of discussion below the average (\textless1.23\%). This might be an indication of a general lack of knowledge, prioritisation, or consideration within these programming languages. 

Using one way analysis of variance (ANOVA) \citep{howell2012} we confirm that there is no significant correlation between the rate of security discussion and the type class, compilation class or memory class paradigms of a programming language. We also confirm that there is no significant correlation between the creation date of a programming language and its rate of security discussion, using the Kendall Tau Coefficient \citep{knight1966}. 

\begin{tcolorbox}
    \textbf{RQ1 Summary:} The average frequency of security discussion is around 1\% for both Stack Overflow and GitHub. Shell-based and Web-oriented languages exhibit the highest rates of discussion (2.6\% and 2.1\% for Shell and PHP respectively), whereas scientific programming-based languages exhibit the lowest (0.1\% for R, Julia and MATLAB). 
\end{tcolorbox}

\subsection{RQ2: What is the intention behind security discussions for different programming languages?}

\begin{figure*}[h]
    \centering
    \includegraphics[width=\textwidth]{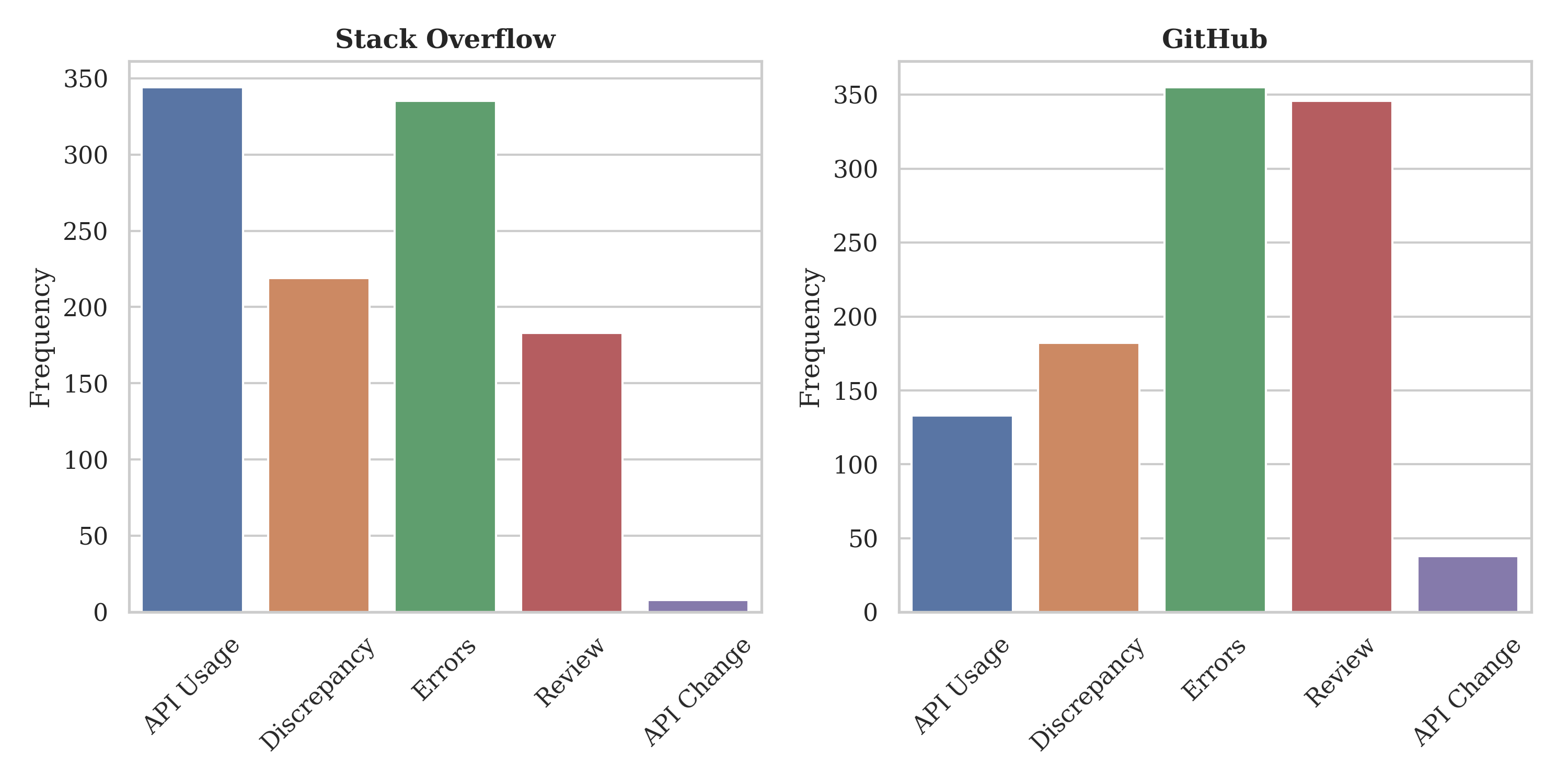}
    \caption{The overall discussion categories for each source }
    \label{fig:overall_lang_purpose}
\end{figure*}

\begin{figure*}[h]
    \centering
    \includegraphics[width=\textwidth]{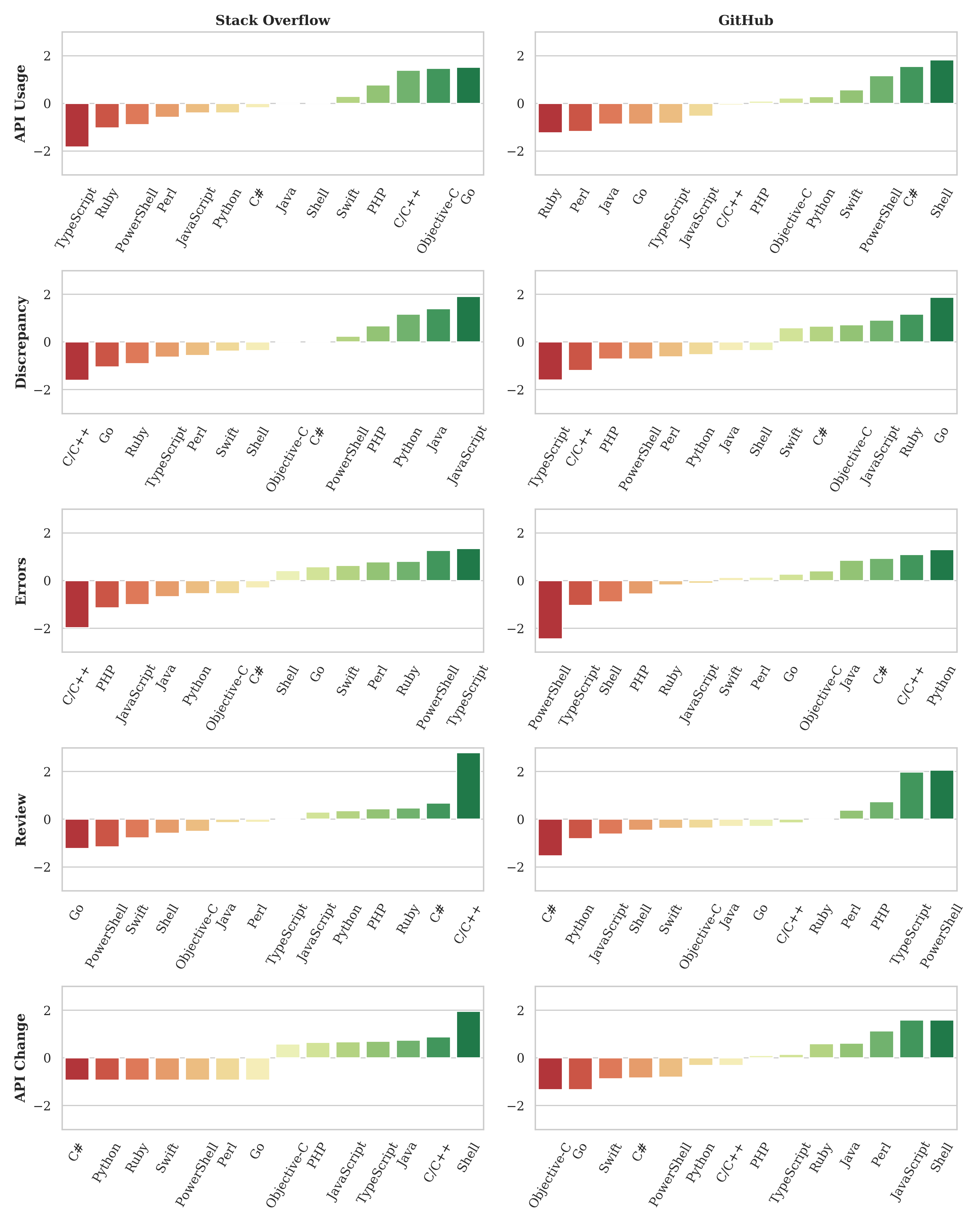}
    \caption{The variance in standard deviation of the distribution of discussion categories for each of the selected programming languages }
    \label{fig:lang_purpose}
\end{figure*}

Using a manually labelled sample, we examine the types of posts using the question taxonomy of \cite{beyer2020}, to determine the purpose and intent of the security discussions for each language. The manually labelled and annotated samples as well as the full category distributions can be viewed in our online appendix\footnote{\url{https://github.com/RolandCroft/Language-Security-Challenges/tree/master/Manual_Analysis}}. 

For the seven proposed question categories, we find that the \textit{Learning} category was extremely infrequent in our dataset. Only 3 out of 952 posts were assigned this label for Stack Overflow, and none for GitHub, so we exclude this category from our analysis. We found the nature of the discussions in our dataset to be more functionally orientated. Similarly, the \textit{Conceptual} category was also under-represented. As most posts discussed implementation related tasks (especially for GitHub), we found the separation between this category and \textit{Review} to be indistinct. We have hence merged the two categories to increase label consistency. 

We present the overall trends for each category in Fig. \ref{fig:overall_lang_purpose}. For Stack Overflow, \textit{API Usage} and \textit{Errors} are the most prominent reasons for discussion as users are commonly asking for assistance in solving a problem. For GitHub, there is a much higher proportion of \textit{Review} questions, as users require approval for proposed changes. We next inspect the difference in distribution for individual programming languages. 

We first consider whether there is a difference in the distribution of question categories for different programming languages. Using a Chi-Squared Test of Independence \citep{pearson1900}, we find that there is an association between the programming language and the category of a post, with p \textless 0.01 for both Stack Overflow and GitHub. Using this knowledge, we then examine how the distribution of question/issue types varies to the standard distribution of each category. Fig. \ref{fig:lang_purpose} displays the standard score (difference in standard deviations to the mean) for the distribution of each question/issue category for each language. Languages with green bars consist of a higher than average proportion of posts with that type of question/issue, and a lower proportion for red bars. 

Through these question/issue types, we can infer the motivations and purpose of the security discussions for each language, and consequently identify the potential deficiencies: 
\begin{itemize}
    \item \textit{API Usage:} This category relates to the posts discussing implementation or use of a specific API. We find that posts of this nature indicate that security is generally hard for a particular language. Implementation is intensive or lacking documentation; hence users need support through discussion. For instance, \textit{``how to add interceptor in spring'' (SO, 8033100)} or \textit{``clarification of documentation for KeePassXC entry protection`` (GH, asbru-cm/asbru-cm/issues/565)}. 
    
    \item \textit{Discrepancy:} This category involves the posts where users encounter unexpected behaviour. The user wants to know what or why something does not work. We observe that the posts of this nature imply a lack of familiarity with security flaws and issues. Perhaps due to a lack of experience, the users have little understanding or knowledge of the security concept or functionality, which may lead to confusion. For instance, \textit{``PHP Login script only works with usernames known to mysql?'' (SO, 56138399).} 
    
    \item \textit{Errors:} These posts are similar to those of the \textit{Discrepancy} category, but explicitly involve exceptions or errors. For GitHub, this category is extended to also consider issues describing how to fix these errors. This category implies that implementation of security is difficult or error-prone. Users make mistakes or miss critical functionality, which needs to be fixed. For example, \textit{``\,`Unauthorized' error in .net webservices developed in C\#'' (SO, 1848508)}. 
    
    \item \textit{Review:} This category is merged with the \textit{Conceptual} category.  These posts attempt to review or understand concepts and functionality, for the purpose of improvement or decision making. For GitHub, this is extended to include code improvement and updates, as users are seeking review and validation. Questions of this nature may indicate that a user's current implementation and security expertise are not sufficient to meet security requirements; they are not confident enough in their secure coding skills to ensure security. For instance, \textit{``Is this sufficient security for user input in PHP?'' (SO, 11003851)}. 
    
    \item \textit{API Change:} Posts for this category discuss compatibility issues and discrepancies between API and software versions. This implies that security issues arise from external software, either from package vulnerabilities or deprecated security features. For example, \textit{``Chrome drops support for added credentials'' (GH, Codeception/Codeception/issues/4384)}, or \textit{``DeprecationWarning: crypto.DEFAULT\_ENCODING is deprecated'' (GH, textlint/textlint/issues/620)}. 
    
\end{itemize}

Security discussions for C/C++ are more heavily oriented towards \textit{Review} than other languages. We observe that developers are much more conscientious of security for these languages. Questions often involve explanation of a vulnerability or attack (e.g., \textit{``Why check if pointer is NULL after using it?'' (SO, 26969565)}, \textit{``How to use Format String Attack?'' (SO, 27018864)}), improvement of code (e.g., \textit{``In C, what is a safer function to use than strtrns?'' (SO, 9264942)}), or security confirmation (e.g., \textit{``Do the openssl X509\_verify\_cert() verifies the signature in the certificate?'' (SO, 10495903)}). This implies that developers are aware of the secure coding requirements for C and C++ \citep{seacord2005}, but require knowledge support. To account for this variation, there is a lower than average distribution of questions for \textit{Errors} and \textit{Discrepancy}. 

PowerShell issues are less oriented towards \textit{Errors} than other languages. They are also more oriented towards \textit{Review}, but neither of these trends are reflected in Stack Overflow questions. Upon inspection, we found that a large proportion of GitHub issues related to PowerShell involved updates or feedback on documentation. This is due to the inclusion of Microsoft Documentation repositories\footnote{https://github.com/MicrosoftDocs}, which are written in PowerShell, but primarily oriented to documentation. 

TypeScript also has a high distribution of issues about \textit{Review}, as discussions were most commonly about security code/feature updates and improvement, e.g., \textit{``Improve the settings sync flows with authentication'' (GH, microsoft/vscode/issues/94766)}. This indicates that security in TypeScript is intensive and requires a lot of implementation. 

Shell discussions are more oriented towards \textit{API Change} on both Stack Overflow and GitHub. We find that developers typically use Shell scripts to install, call or update security related packages and components. Hence, these posts typically discuss version issues (e.g., \textit{``Issue with Debian 10 \textless tls-crypt \textgreater empty'' (GH, angristan/openvpn-install/issues/523)}) or updates (e.g., \textit{``update oraclelinux for openssl'' (GH, docker-library/official-images/issues/4957)}, \textit{``Update the OpenVPN mirror GPG key and fingerprint'' (GH, StreisandEffect/streisand/issues/754)}). 

\begin{tcolorbox}
    \textbf{RQ2 Summary:} There is a correlation between programming language and the purpose of the security discussions. C/C++ developers are much more conscientious of security but require knowledge support. TypeScript issues are often about security updates and improvements. Security challenges for Shell are motivated through use of external components. 
\end{tcolorbox}

\subsection{RQ3: What are the major developer security discussion topics?}
\begin{figure*}[h]
    \centering
    \includegraphics[width=\textwidth]{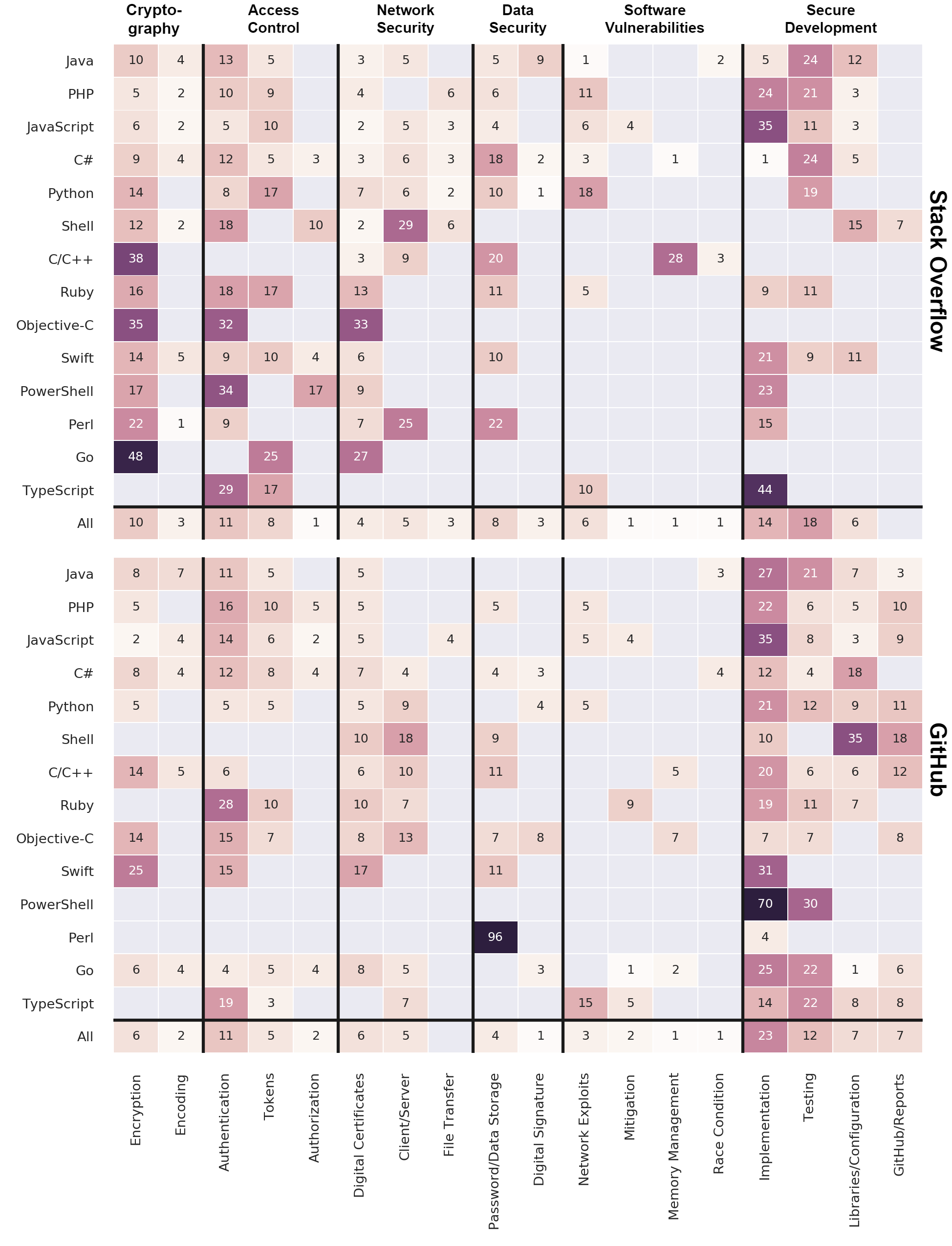}
    \caption{The heatmap of the topic share (\%) for the identified topic sub-categories for the 15 selected programming languages for Stack Overflow and GitHub}
    \label{fig:topic_heatmap}
\end{figure*}

To address this research question, we first create a taxonomy of the security discussion topics, as identified via the topic models. The taxonomy is displayed in Fig. \ref{fig:topic_heatmap}. This taxonomy is based on the dominant and emergent discussion topics in our dataset. It should be noted that this taxonomy is not exhaustive, nor does this encompass all elements of cybersecurity. The taxonomy is focused around the content of our dataset, security related development issues, rather than concepts and knowledge. Furthermore, if a category is absent for a particular language, it does not necessarily mean that the discussion is non-existent; it may just be overshadowed by other topics. 

Labels are not assigned to topics which do not implicitly relate to security or are too ambiguous to label. Eleven out of the 395 total individual topics of all languages and sources were not assigned a label, and thus excluded from this study. The impact of this exclusion is minimal as the 11 topics each have a topic share value of less than 1. 

The heatmap of the topic distribution for each programming language of each source is shown in Fig. \ref{fig:topic_heatmap}, where each value is the percentage topic share. Cells are rounded to nearest integer, where cells with no number have a topic share value rounded to 0. As such, rows do not necessarily total 100. It should be noted that the distribution of each programming language is limited to topics which actually appeared in the topic models. Thus, languages with less diversity tend to have higher topic share values. The overall topic share value for each data source (considering all documents of a data source as a single corpus, regardless of language) is given in the \textit{All} row to give an indication of the overall topic distribution. An explanation and analysis of the taxonomy categories is below. 

\textbf{Category - \textit{Cryptography}:} Cryptography relates to the techniques and practices for secure communication and information security. Cryptographic systems are a security primitive, essential for ensuring data confidentiality, data integrity, authentication and non-repudiation \citep{menezes2018}. As such, this category exhibits some overlap with other categories, but the distinction lies in the focus of the techniques and methods, rather than the application and uses. Noticeably, TypeScript exhibits no Cryptography related discussions or challenges on both sources. 

\begin{itemize}
    \item \textbf{Encryption:} All discussion relating to encryption/decryption and relevant cryptographic algorithms, without any explicit application. For instance, \textit{``How do you encrypt large files / byte streams in Go?'' (SO, 49546567)} or \textit{``Encrypted String is wrong in swift'' (GH, krzyzanowskim/CryptoSwift/issues/774)}. This is one of the more prominent sub-categories for both Stack Overflow and GitHub, due to its diverse applications. Across the two sources, discussion is most dominant for C/C++, Objective-C, Swift and Go, with each of these languages exhibiting a quarter of their overall discussion for this topic. Topics for C/C++, Shell, JavaScript and Java have a bigger focus on the cryptographic algorithms rather than encryption itself (e.g., \textit{``Deriving ECDSA Public Key from Private Key'' (SO, 49204787)}). This is most noticable for C/C++ due to the large number of discussions for \textit{OpenSSL}\footnote{https://www.openssl.org/}, a cryptography toolkit. This indicates that use of cryptography is very strong for this language. 
    
    \item \textbf{Encoding:} The encoding or decoding of data. Whilst encoding is not explicitly related to cryptography, it was found that the majority of discussions of this topic fell under the cryptography category. For instance, the post \textit{``Could not parse base64 DER-encoded ASN.1 public key from iOS in Golang'' (SO, 48761010)} discusses encoding a public key in Go. 
\end{itemize}

\textbf{Category - \textit{Access Control}:} Access Control involves the methods, practices and techniques of restriction and permissive access to resources, and other access management processes. This is the second most prominent category, behind Secure Development. Challenges occur for all languages across either source. 

\begin{itemize}
    \item \textbf{Authentication:} Discussion relating to user validation, typically through login, accounts or credentials. We found the majority of topics and posts for this sub-category to relate to implementing a login system, e.g., \textit{``Ruby on Rails: Devise - password change on first login'' (SO, 13121356)}. As such, challenges are particularly prominent in languages oriented for app development (i.e., Web or Mobile development), like PHP, C\#, Ruby, Objective-C, Swift, PowerShell, and TypeScript. The high overall topic share value for these discussions suggest that it is one of the most common security-related tasks that developers face. 

    \item \textbf{Tokens:} Token-based authentication and other similar mechanisms, such as API tokens, cookies, and general session management. This sub-category is more frequent on Stack Overflow, for languages such as Python, Ruby, Go and TypeScript. API tokens such as OAuth token authentication are a prominent discussion point, e.g., \textit{``Google oauth how Exchange code for access token and ID token'' (SO, 22796143)}. 

    \item \textbf{Authorization:} The process of assigning and checking user permissions to access a resource. This category is noticeably significant on Stack Overflow for Shell (10\%) and PowerShell (17\%) due to the use of methods such as Secure Shell (SSH) for connecting to remote servers (i.e., \textit{``Validate SSH connectivity using SSH-keys'' (SO, 38006170)}). For all languages, discussion of Authorization is less than that of other Access Control sub-categories. 
\end{itemize}

\textbf{Category - \textit{Network Security}:} Network Security involves the practices required to enable secure communication and data transfer over a computer network. 

\begin{itemize}
    \item \textbf{Digital Certificate:} Secure communication through the use of digital certificates. One of the more prominent topics for Network Security, with the highest topic share values for Objective-C (33\%) and Swift (17\%) on Stack Overflow and GitHub respectively. These languages are both commonly used for iOS development, e.g., \textit{``How do I accept a self-signed SSL certificate using iOS 7's NSURLSession'' (SO, 30739149)}. 

    \item \textbf{Client/Server:} Secure communication over a network to a server and/or client. This sub-category is most prominent for Shell (29\% and 10\% Topic Share for Stack Overflow and GitHub respectively), again due to the use of Secure Shell (SSH). For example, \textit{``How do I access a WebService through an SSH tunnel?'' (SO, 2289708)}. Besides Shell, this topic is also noticeably evident for Perl and C/C++. These languages are commonly utilized for socket programming. For instance, this issue \textit{``net.socket:connect() - leaks memory on DNS lookup fail'' (GH, nodemcu/nodemcu-firmware/issues/234)} is about a security problem for a socket connection in C. 

    \item \textbf{File Transfer:} Secure transfer and access of files over a network, e.g., \textit{``Security threats with uploads'' (SO, 11061355)}. This topic is most prominent for PHP and JavaScript on Stack Overflow and GitHub respectively, but it is overall relatively infrequent. 
\end{itemize}

\textbf{Category - \textit{Data Security}:} Data Security involves protecting digital data from attacks or breaches. 

\begin{itemize}
    \item \textbf{Password/Data Storage:} Secure storage of passwords or data, usually through cryptographic means of encryption or hashing. Discussions for this topic often relate to implementation of secure password stores, e.g., \textit{``Hash and salt passwords in C\#'' (SO, 2138429)}. As such, this topic is noticeably evident in languages commonly used for app development.     Interestingly, almost all of the security-related documents on GitHub for Perl (96\%) are for this topic, as well as a large portion from Stack Overflow (22\%).  Aside from the higher proportion, we did not observe any distinct characteristics of the Perl posts in comparison to other languages. 

    \item \textbf{Digital Signature:} Verifying the authenticity of data or documents through the use of digital signatures, e.g. \textit{``Add script to verify signature'' (GH, sparkle-project/Sparkle/issues/896)}. Due to the much higher topic share value for Password/Data Storage, it implies that most challenges in data security relate to confidentiality rather than integrity or authenticity. 
\end{itemize}

\textbf{Category - \textit{Software Vulnerabilities}:} This category relates to security weaknesses present in a software system that can be exploited by an attacker. This involves the explicit flaws in the code or security features. 

\begin{itemize}
    \item \textbf{Network Exploits:} Network exploits and vulnerabilities, such as SQL Injection, XSS, CSRF and DDoS, as well as methods to prevent them, such as input validation. For example, \textit{``How can prepared statements protect from SQL injection attacks?'' (SO, 8263371)}. As these attacks are often targeted at web applications, this topic is noticeably most prominent in web-oriented languages, i.e., PHP, JavaScript, Ruby and TypeScript. However, challenges are most frequent for Python on Stack Overflow, with 18\% Topic Share. This is because of Python's \emph{Django} framework enforcing CSRF protection, leading to many Stack Overflow questions. For example, \textit{``When I try to send POST request, Django requires CSRF token. How can I send POST request without any problems?'' (SO, 23743273).} 
    
    \item \textbf{Mitigation:} Vulnerability reports, scanning and patches. This topic does not fall under other sub-categories as the discussion is often not explicit to any particular vulnerability. For instance, the GitHub issue \textit{``chore: fix vulnerability'' (GH, strongloop/loopback-next/pull/2880)} does not provide details about the actual vulnerability it fixes. Like Network Exploits, this topic is evident for web-oriented languages: JavaScript, Ruby and TypeScript. 
    
    \item \textbf{Memory Management:} Discussion relating to the management and mitigation of memory related issues. The dominant challenges in this topic relate to stack/heap memory and overflows, e.g., \textit{``buffer overflow exploits - Why is the shellcode put before the return address'' (SO, 16789241)}. Individual topics can also relate to dynamic memory linking methods, such as pointers, but this discussion is only prominent in C and C\# as these mechanisms are often uncommon in other languages. GitHub issues for this topic commonly relate to memory leaks, e.g., \textit{``Fixes memory leaks in IoT'' (GH, aws-amplify/aws-sdk-ios/pull/1175)}. This topic is only dominant for C/C++, and discussion is noticeably lacking for other languages. This is potentially expected, as these languages are memory and type unsafe, but these vulnerabilities can still manifest in other programming languages, such as PHP \citep{cifuentes2019}. 

    \item \textbf{Race Condition:} Secure and safe management of access to shared resources by multithreaded or parallel processes, e.g., \textit{``Can Multiple Threads Read The Same Class Member Variable?'' (SO, 59191339)}. This is the least prominent sub-category, and is only evident for Java, C\#, and C/C++. 
\end{itemize}

\textbf{Category - \textit{Secure Development}:} Secure development refers to the practices and tasks involved with producing secure software. These discussions do not hold any explicit relation to the other categories. For example, debugging a security feature, or configuring a secure framework. This category has the highest overall topic share value, as the major focus of the collected dataset is secure development. As a result, discussion is prominent for all languages in our dataset. 

\begin{itemize}
    \item \textbf{Implementation:} Implementation or requirements planning of security elements, e.g., \textit{``Submit button is not getting enabled after checking all the fields'' (SO, 34253518)} or \textit{``Add best practice encryption requirements.'' (GH, coreinfrastructure/best-practices-badge/pull/3)}. 
    
    \item \textbf{Testing:} Testing and debugging for general security defects or bugs, e.g., \textit{``initializationError FAILED java.lang.SecurityException'' (SO, 29508622)}. 

    \item \textbf{Libraries/Configuration:} Integration and configuration of security-related libraries, frameworks, plugins or addons; e.g., \textit{``spring security 3.2 java configuration'' (SO, 26944600)}. 

    \item \textbf{GitHub/Reports:} General GitHub development activities and reporting of software issues and bugs, such as GitHub issues and pull requests, or bug reports. For GitHub, this topic is heavily influenced by GitHub bots, e.g., \emph{googlebot}\footnote{\url{https://developers.google.com/open-source/github/accounts\#googlebot}} which requests the user to sign a license agreement. This topic is not very pervasive on Stack Overflow as expected, but still appears for Shell due to the frequent usage of Git via the command line, e.g., \textit{``How to switch from first github account to second?'' (SO, 37782617)}.  
\end{itemize}

\begin{figure}[h]
    \centering
    \includegraphics[width=\textwidth]{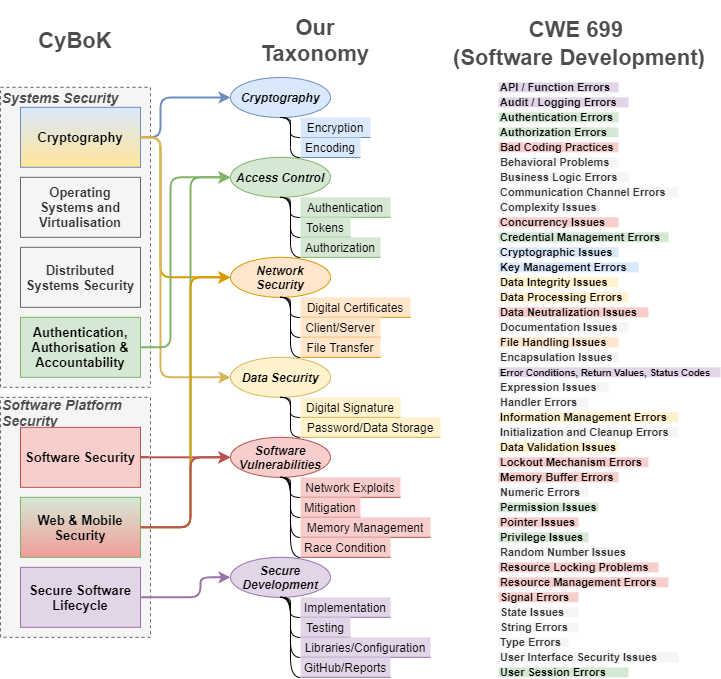}
    \caption{The comparison of our discovered security taxonomy to the security state of the art }
    \label{fig:cat_comparison}
\end{figure}

The existing security state of the art also provides taxonomies for security theory and knowledge. However, none of these taxonomies are empirically constructed through development challenges or mapped to programming languages like ours. We position our findings with respect to the state of the art by comparing our identified taxonomy to two existing security taxonomies. The Cybersecurity Body Of Knowledge (CyBOK) \citep{cybok}, is a mapping and collection of cybersecurity knowledge published in textbooks, academic research articles and technical reports. It identifies 20 knowledgebases for five key domains. Additionally, the Common Weakness Enumeration (CWE) presents a taxonomy of security weaknesses for software security (CWE 699)\footnote{\url{https://cwe.mitre.org/data/definitions/699.html}}. We compare our taxonomy to these existing security taxonomies in Fig. \ref{fig:cat_comparison}. 

As our taxonomy is constructed empirically from development issues rather than security theory, we notice several differences to the existing taxonomies. CyBOK is oriented towards education, and as a result covers security aspects very broadly. Only two of the domains, \textit{System Security} and \textit{Software Platform Security}, relate to the development challenges that we identify. Furthermore, the sub-categories of the relevant chapters do not match the granularity of our analysis and are often topics that are quite specific to theoretical aspects, such as \textit{Cryptographic Security Models}. Additionally, the existing state of the art is not provided from a programming language perspective, which can make application difficult for practitioners, managers, educators, and researchers who are concerned about this view.

The CWE taxonomy is primarily oriented towards vulnerabilities and exploits, which only covers a small portion of our dataset. Challenges often relate to development and implementation of security features rather than discussion of explicit flaws. Some of the more simplistic vulnerabilities are not covered in our taxonomies, such as Numeric, String and Type Errors. Although these vulnerabilities may still receive some discussion, they do not experience enough discussion in developer communities to manifest as significant challenges through our topic modelling process. 

Although, CWE entries are substantiated using references to formal public reports and vulnerability databases, these references are not systematic nor evaluated with respect to publicly available security discussions and sources, like our taxonomy is. Additionally, CWE provides the applicable platforms/programming languages for some CWE types, but this information is considered far from complete. For instance, 75\% of CWE types which list Applicable Platforms (423 out of 566) are categorized as \textit{Language-Independent}, which may be theoretically true, but through our study we find the prevalence and nature of the security challenges to vary across programming language. Furthermore, 63\% (362 out of 571) of the language examples on CWE are for Java and C, which does not provide a wide view of different programming languages. 

\begin{tcolorbox}
    \textbf{RQ3 Summary:} We identify 18 topics and 6 topic categories: Cryptography, Access Control, Network Security, Data Security, Software Vulnerabilities, and Secure Development. Secure Development is the overall most prominent topic category, followed by Access Control. Different languages experience different topic distributions. 
\end{tcolorbox}

\subsection{RQ4: How do security discussion topics change over time for different programming languages?}
We examine the topic impact values of each topic category for each language on Stack Overflow and GitHub for the time span of January 2009 to May 2020. Using the Mann-Kendall Trend Test \citep{Hussain2019} with a 99\% significance level, we identify each category as either increasing, decreasing or no-trend. The results can be seen in Table \ref{table:topic_trend}. 

Topic Impact appears to be predominantly increasing for most topic categories, especially amongst languages on GitHub and topics.  Secure Development in particular exhibits a steadily increasing trend across nearly all languages and sources. 

All languages exhibit an overall increasing trend in security challenges, particularly for Secure Development. This implies a general increase in security discussion and challenges over the past decade. PHP is the only language which has a higher number of decreasing topic trends across both sources.

As questions, issues and discussions usually rise in response to problems, an increasing Topic Impact value may imply a rise in challenges. However, an increase in discussion also represents increased awareness and a larger knowledge-base, especially for Stack Overflow. 

\begin{table}[h]
\caption{Overall topic impact trend from January 2009 to May 2020}
\label{table:topic_trend}
\resizebox{\columnwidth}{!}{%
\begin{tabular}{ |c|c|c|c|c|c|c| } 
    \hline
    Programming & \multicolumn{6}{|c|}{Topic Category}\\
    \cline{2-7}
    Language & Cryptography & Access Control & Network Security & Data Security & Software Vulnerabilities & Secure Development\\
    \hline
    Java & \DownArrow  \DownArrow & \UpArrow  \Dash & \DownArrow  \UpArrow & \DownArrow  $\times$ & \UpArrow  \UpArrow & \UpArrow  \UpArrow\\
    \hline
    PHP & \UpArrow  \Dash & \UpArrow  \DownArrow & \Dash  \UpArrow & \DownArrow  $\times$ & \DownArrow  \DownArrow & \DownArrow  \UpArrow\\
    \hline
    JavaScript & \UpArrow  \Dash & \UpArrow  \UpArrow & \DownArrow  \UpArrow & \UpArrow  $\times$ & $\times$  \UpArrow & \DownArrow  \UpArrow\\
    \hline
    C\# & \Dash  \Dash & \UpArrow  \Dash & \DownArrow  \UpArrow & \DownArrow  \UpArrow & \DownArrow  \Dash & \UpArrow  \UpArrow\\
    \hline
    Python & \Dash  \UpArrow & \UpArrow  \DownArrow & \UpArrow  \Dash & \Dash  \Dash & \DownArrow  \UpArrow & \Dash  \UpArrow\\
    \hline
    Shell & \Dash  $\times$ & \UpArrow  $\times$ & \DownArrow  \UpArrow & $\times$  \UpArrow & $\times$  $\times$ & \UpArrow  \UpArrow\\
    \hline
    C & \UpArrow  \UpArrow & $\times$  \DownArrow & \Dash  $\times$ & \DownArrow  $\times$ & \Dash  $\times$ & \Dash  \UpArrow\\
    \hline
    Ruby & \Dash  $\times$ & \Dash  \DownArrow & \Dash  \UpArrow & \DownArrow  $\times$ & \Dash  \UpArrow & \Dash  \Dash\\
    \hline
    Objective-C & \DownArrow  \UpArrow & \Dash  \UpArrow & \UpArrow  \Dash & $\times$  \UpArrow & $\times$ \Dash & $\times$  \UpArrow\\
    \hline
    Swift & \UpArrow  \UpArrow & \UpArrow  \UpArrow & \UpArrow  \UpArrow & \UpArrow  \UpArrow & $\times$  $\times$ & \UpArrow  \UpArrow\\
    \hline
    PowerShell & \UpArrow  $\times$ & \Dash  $\times$ & \UpArrow  $\times$ & $\times$  $\times$ & $\times$  $\times$ & \Dash  \UpArrow\\
    \hline
    Perl & \Dash  $\times$ & \Dash  $\times$ & \Dash  $\times$ & \Dash  \Dash & \UpArrow  $\times$ & \Dash  \UpArrow\\
    \hline
    Go & \UpArrow  \UpArrow & \UpArrow  \DownArrow & \UpArrow  \UpArrow & $\times$  \UpArrow & $\times$  \UpArrow & $\times$  \UpArrow\\
    \hline
    TypeScript & $\times$  $\times$ & \UpArrow  \UpArrow & \UpArrow  \UpArrow & $\times$  $\times$ & $\times$  \UpArrow & \UpArrow  \UpArrow\\
    \hline
\end{tabular}%
}
\tiny \textbf{Note:} The symbol on the left indicates the Stack Overflow trend and the symbol on the right indicates GitHub. An up arrow represents an increasing trend, a down arrow represents a decreasing trend, a dash represents no trend, and a cross represents no topics for that category.
\end{table}

The Access Control topic category is overall decreasing on GitHub, which could be interpreted as errors in Access Control are not occurring as commonly. PHP, Python and Go have an increasing trend for Access Control on Stack Overflow, despite the decreasing trend on GitHub. Access Control challenges are still increasing for JavaScript, Swift and TypeScript on both sites.

For Stack Overflow, there are several decreasing trends for Network Security and Data Security. This implies a growing lack of prevalence on these topics for the relevant languages. These decreasing trends are for the most popular languages of Stack Overflow developers\footnote{\url{https://insights.stackoverflow.com/survey/2020}}. Thus, these languages may be moving to other areas of focus due to their popularity. For Network Security, none of these trends are reflected on GitHub, which implies they are still important factors of development. However for Data Security, this topic is not represented commonly on GitHub. 

These decreasing trends may be further supported by the sites attempted avoidance of duplicate questions. Security questions and discussions that are posted earlier are typically not re-posted, as users have the ability to visit the original post. 

\begin{figure}[h]
    \centering
    \includegraphics[width=\textwidth]{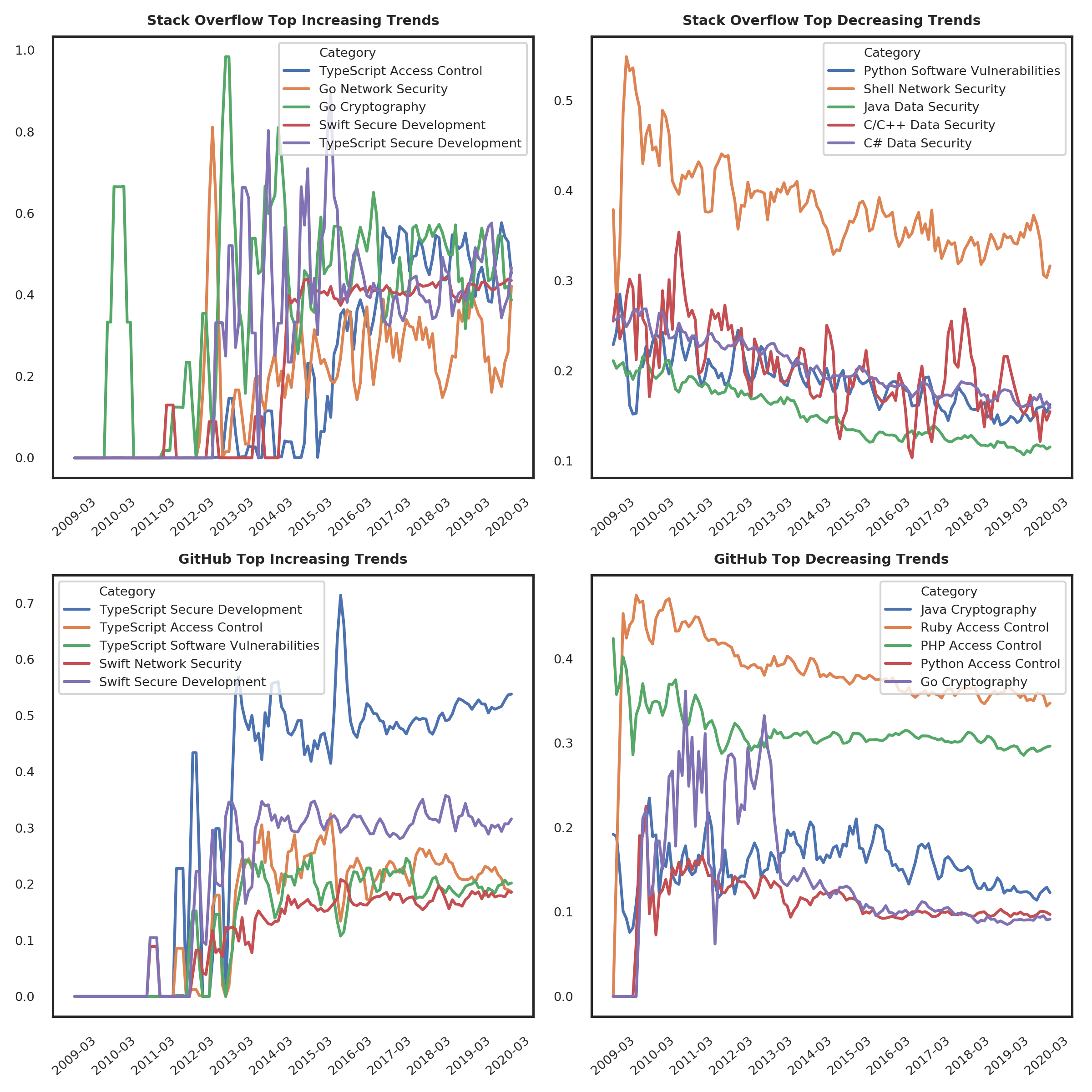}
    \caption{The 5 most significant increasing and decreasing topic category trends for Stack Overflow and GitHub}
    \label{fig:top_trends}
\end{figure}

We examine the five most significant positive and negative trends for both Stack Overflow and GitHub in Fig. \ref{fig:top_trends}, as calculated by the Theil-Sen's Slope estimation value \citep{Hussain2019}. We use a rolling average of window size 3 to help smooth the noise in these trend graphs. 

The major increases in both Stack Overflow and GitHub language security appear to be caused by the inception of new languages. The major spikes in the top left graph of Fig. \ref{fig:top_trends} align to the date that these languages first commonly appeared; Go: 2009, TypeScript: 2012, Swift: 2014. As such, the trends for these graphs are noisy, but we can still observe an initial surge of interest followed by a steady increase in discussion/challenges for the various security topics of these languages, after their original inception. 

This trend may be caused by general inexperience with these newer languages or lacking documentation. For example, \textit{``I am new to typescript. I am able to integrate auth0 using javascript because auth0 providing a sample for that but there is no sample app available for vue with typescript.'' (SO, 59135589)}. This user admits that they are a novice to TypeScript and are struggling to find support for this language. 

\begin{tcolorbox}
    \textbf{RQ4 Summary:} The most popular languages experience decreasing challenges in the areas of: Access Control (GitHub), Network Security (Stack Overflow), and Data Security (Stack Overflow). However, challenges are overall increasing for most languages, except for PHP. The most significant trends are for new languages, which see a steady rise in challenges after their introduction. 
\end{tcolorbox}

\subsection{RQ5: What are the characteristics in terms of popularity and difficulty of different programming languages and their identified security challenges?}
\begin{figure}[h]
    \centering
    \includegraphics[width=0.9\textwidth]{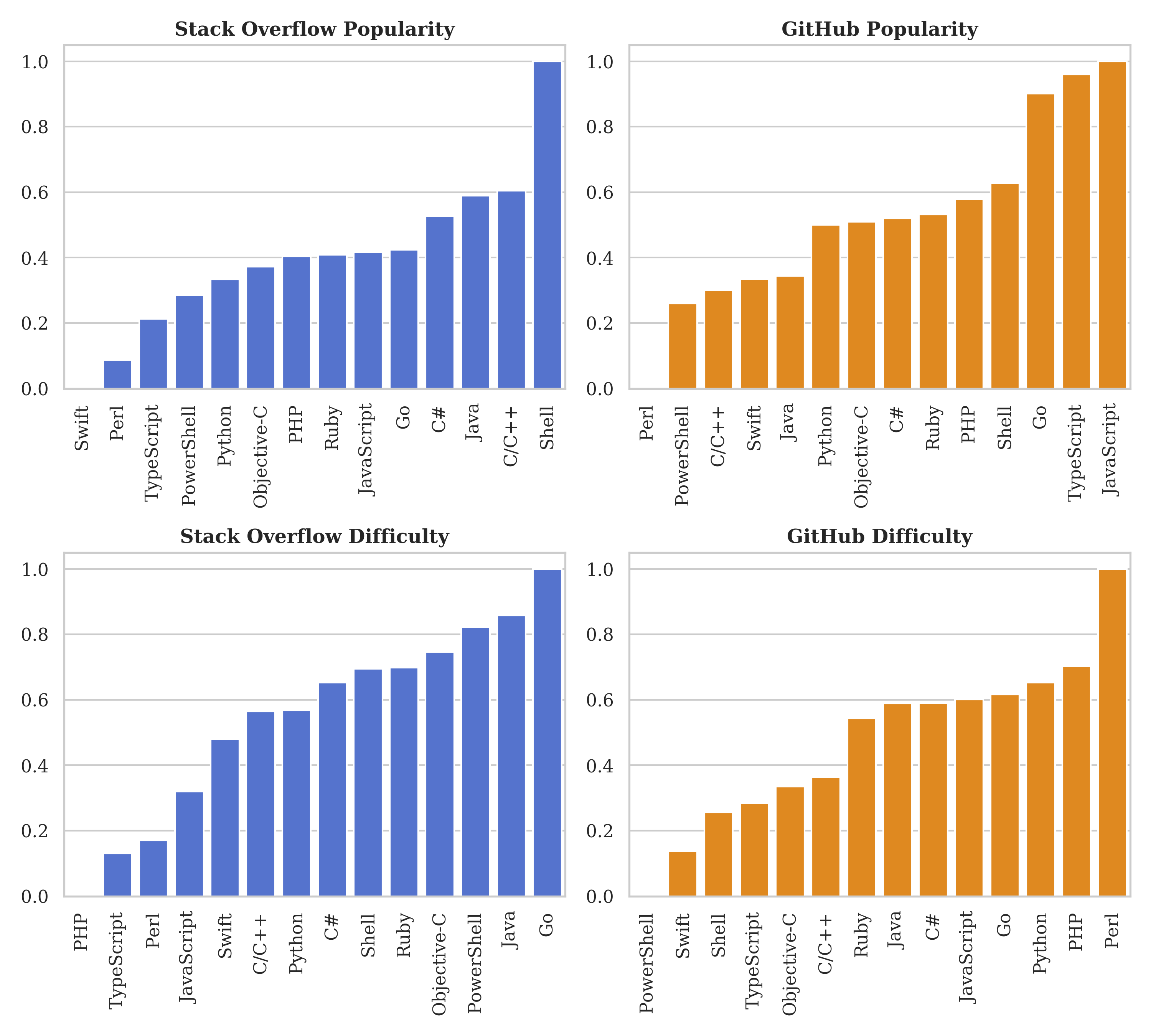}
    \caption{The normalized popularity and difficulty values of each language for Stack Overflow and GitHub}
    \label{fig:overall_pop_diff}
\end{figure}

The popularity and difficulty metrics provide insights into the characteristics of the identified and analyzed security challenges. Through these insights we can better inform developers of the nature of these challenges; their ability to be resolved and handled. The popularity metric gives an indication of the interest and involvement of users for these challenges, which suggests how readily these challenges will be solved by online communities. The difficulty metric gives an indication of the complexity and hardness of the challenges; how much knowledge and labour is required to overcome the identified issues. The combination of these metrics gives an indication of how significant the challenges are when they occur. High popularity and low difficulty challenges are preferable as they would be expected to be resolved relative quickly and easily. 

The overall normalized popularity and difficulty for security of the languages for each source is presented in Fig. \ref{fig:overall_pop_diff}. It is expected that developers would favor languages which are both popular and not difficult; as it implies that there is an abundant amount of interest and resources to solve potential issues, and that the encountered problems are not as challenging. For instance, \cite{meyerovich13} find language popularity and learning cost to be important factors for language adoption. 

\begin{figure}[h]
    \centering
    \includegraphics[width=0.75\textwidth]{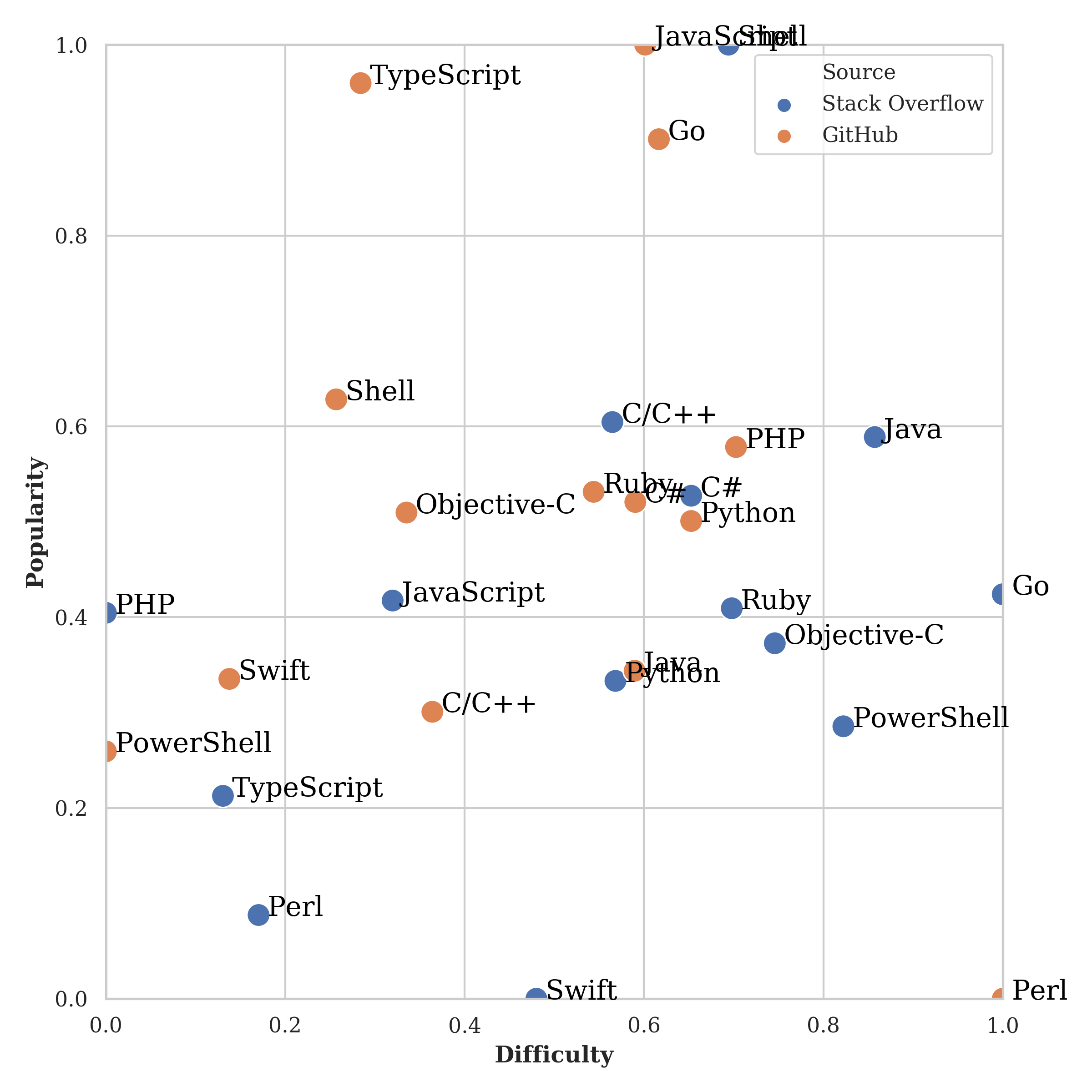}
    \caption{The normalized popularity against the normalized difficulty values of each language. Sources are normalized separately }
    \label{fig:pop_diff_categories}
\end{figure}

We observe that there is no significant correlation between the Stack Overflow and GitHub metric values, using Kendall's Rank Correlation Coefficient \citep{knight1966}. Similarly, there is no significant correlation between the Popularity and Difficulty values, which implies that the most popular languages and topics for security are not necessarily the least difficult and vice versa. Fig. \ref{fig:pop_diff_categories} displays the normalized popularity against difficulty for each language and source. 

On Stack Overflow, security related posts for Shell are significantly the most popular. Upon closer inspection of the individual metrics, this is because Shell related security posts receive significantly more views from the community. We find that high view posts for Shell commonly relate to SSH Keys, an authentication method for an ecrypted connection between systems. For instance, \textit{``Automatically enter SSH password with script''} or \textit{``Calculate RSA key fingerprint''}. The popularity of these posts is likely due to the diverse usage of SSH keys regardless of the application context. C/C++, Java and C\# are also popular on Stack Overflow. 

The most popular languages on GitHub for security are JavaScript, TypeScript and Go. The popularity of JavaScript and TypeScript on GitHub implies that security is a very popular topic for web development, receiving active interest from the community. Security related posts for web development languages (i.e., PHP, TypeScript, JavaScript) are also the least difficult on Stack Overflow. 

Perl appears to be an undesirable language amongst the community for security. Posts receive very little attention (Popularity) from both the Stack Overflow and GitHub community. It also has the most difficult to resolve issues on GitHub. 

\begin{table*}[h]
\caption{The most and least Popular and Difficult topic categories for each language}
\label{table:mostleast_metrics}
\resizebox{\textwidth}{!}{%
\begin{tabular}{ |p{1cm}|p{2cm}|p{2cm}|p{2cm}|p{2cm}|p{2cm}|p{2cm}| }
    \hline
    \multicolumn{7}{|c|}{\textbf{Stack Overflow}}\\
    \hline
    Metric & Cryptography & Access Control & Network Security & Data Security & Software Vulnerabilities & Secure Development\\
    \hline
    \hline
    Most Popular & \textbf{\textit{Shell}}, Objective-C, \textbf{\textit{Perl}} & - & PowerShell, \textbf{\textit{Go}} & Java, PHP, C\#, Ruby, Swift & Python, \textbf{\textit{C}}, \textbf{\textit{TypeScript}} & JavaScript \\
    \hline
    Least Difficult & Python, \textbf{\textit{Shell}}, \textbf{\textit{Perl}} & Objective-C, PowerShell & \textbf{\textit{Go}} & JavaScript & Java, PHP, C\#, \textbf{\textit{C}}, Ruby, \textbf{\textit{TypeScript}} & Swift \\
    \hline
    \hline
    Least Popular & Ruby & Java, PHP, Python, Objective-C, PowerShell, Perl, \textbf{\textit{Go}}, \textbf{\textit{TypeScript}} & \textbf{\textit{Shell}}, \textbf{\textit{C}} & JavaScript & C\# & Swift \\
    \hline
    Most Difficult & PHP, JavaScript & \textbf{\textit{Go}}, \textbf{\textit{TypeScript}} & Python, \textbf{\textit{Shell}}, \textbf{\textit{C}}, Ruby, Objective-C, PowerShell & Swift & - & Java, C\#, Perl \\
    \hline
    
    \multicolumn{7}{c}{\vspace{0.25cm}}\\
    
    \hline
    \multicolumn{7}{|c|}{\textbf{GitHub}}\\
    \hline
    Metric & Cryptography & Access Control & Network Security & Data Security & Software Vulnerabilities & Secure Development\\
    \hline
    \hline
    Most Popular & C\# & - & JavaScript, Python, Ruby, \textbf{\textit{TypeScript}} & \textbf{\textit{PHP}}, Shell, C, \textbf{\textit{Objective-C}}, Swift, Perl & - & Java, \textbf{\textit{PowerShell}}, Go \\
    \hline
    Least Difficult & Swift & C\# & \textbf{\textit{TypeScript}} & \textbf{\textit{PHP}}, Python, \textbf{\textit{Objective-C}} & Java, JavaScript, C, Go & Shell, Ruby, \textbf{\textit{PowerShell}}, Perl \\
    \hline
    \hline
    Least Popular & JavaScript, Python & Ruby & Shell, Swift & - & Java, PHP, C, \textbf{\textit{Objective-C}}, Go, TypeScript & C\#, \textbf{\textit{PowerShell}}, Perl \\
    \hline
    Most Difficult & C\#, C & PHP, Swift, Go, TypeScript & JavaScript, Ruby & Shell, Perl & \textbf{\textit{Objective-C}} & Java, Python,  \textbf{\textit{PowerShell}} \\
    \hline
\end{tabular}%
}
\end{table*}

We further summarize the categories for which each language is considered most and least Popular/Difficult in Table \ref{table:mostleast_metrics}. The most popular categories for a language receive the most user attention, and the least difficult categories are resolved the fastest and most easily. Hence, from Table \ref{table:mostleast_metrics} practitioners can infer which security aspects of a language receive the most and least community support. Languages which have an inverse ranking for popularity and difficulty are highlighted in bold. 

The individual values for the metrics that comprise the Popularity and Difficulty Score of each topic for each language can be seen in our online appendix\footnote{\url{https://github.com/RolandCroft/Language-Security-Challenges/tree/master/Document\_Metrics}}. 

\begin{tcolorbox}
    \textbf{RQ5 Summary:} Popularity and Difficulty differs heavily for languages across Stack Overflow and GitHub. Web-Oriented languages are the most popular on GitHub, whereas they are the least difficult on Stack Overflow. Similarly, different languages exhibit different rankings for individual topic categories. 
\end{tcolorbox}

\subsection{RQ6: What is the level of security expertise of the users who answer security related discussions for different programming languages?}
\begin{figure}[h]
    \centering
    \includegraphics[width=0.8\textwidth]{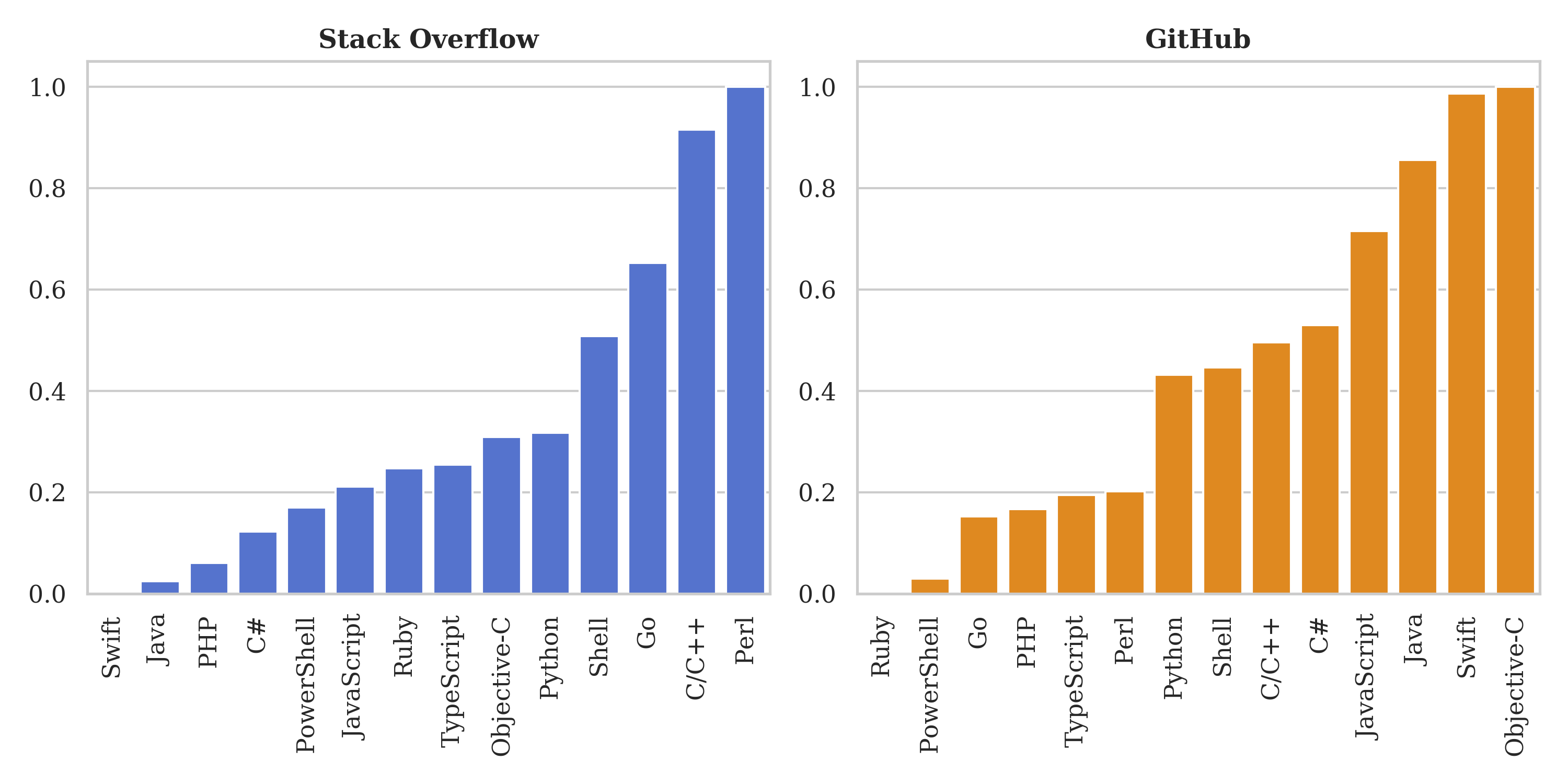}
    \caption{The normalized mean security expertise values of each language for Stack Overflow and GitHub}
    \label{fig:overall_expertise}
\end{figure}

The level of security expertise gives an indication of a user's knowledge and experience for security related topics. For Stack Overflow users, a high expertise value implies a wealth of knowledge, either theoretical or practical, and the ability to convey this knowledge well to other users. For GitHub users, a high expertise value indicates a lot of experience in software development and practices for high quality projects. Through this we can infer whether there are enough experts to resolve the security challenges that a developer may face for a particular programming language, and the quality of the provided answers. A language with high expertise users is considered preferable, as it would be expected that answers and hence resolutions are of higher quality.  This research question helps us better understand how the challenges are resolved. The normalized mean security expertise values for all users who answer security related questions in our dataset is presented in Fig. \ref{fig:overall_expertise}. 

Similar to RQ5, we again observe a disconnect between Stack Overflow and GitHub. For GitHub, the languages with the highest security expertise are languages commonly used for mobile development (i.e., Objective-C, Swift, Java, JavaScript, C\#). However, on Stack Overflow the expertise of users for these languages is relatively low. This might suggest that their is an abundance of experienced secure mobile developers, but their ability to detail knowledge is lacking. 

For Stack Overflow, languages with a lower level of abstraction (Shell, C/C++) have relatively higher expertise values. These languages are typically considered both type and memory unsafe which implies a greater level of security knowledge is required to avoid potential vulnerabilities. Hence, it is a promising sign that the average security expertise of users' answers is relatively high. These languages also rank highly on GitHub, but still have a much lower expertise value than the aforementioned mobile-related languages. Go also ranks highly for Stack Overflow. Whilst this language is syntactically similar to C and C++, it has added memory safety and garbage collection capabilities. 

Perl ranks as the language with the highest average security expertise for Stack Overflow users, which may explain its low difficulty values amongst this community in RQ5. However, the security expertise of Perl users on GitHub is relatively low, which is potentially expected as Perl is the least popular and most difficult language for GitHub.

Except for JavaScript on GitHub, web-oriented languages (i.e., PHP, Ruby, TypeScript, JavaScript) have a relatively low average user security expertise on both sources. This implies that the answers given to challenges for these languages are fairly simplistic. 

\begin{tcolorbox}
    \textbf{RQ6 Summary:} Language security expertise heavily differs for users of Stack Overflow and GitHub. Mobile development oriented languages have the highest expertise users on GitHub, whereas languages with lower levels of abstraction have the highest expertise users on Stack Overflow. 
\end{tcolorbox}

\section{Discussion}
The goal of this study is to empirically investigate and compare developer security challenges across different languages. To achieve this, we investigated 6 research questions to identify the different programming language security consideration, discussions and support. 

We observe that security related topics are relatively infrequent, only forming an average of 1.23\% of all discussion for programming languages. In a recent developer survey by \cite{venson2019}, it was estimated that security accounts for an average of 20\% of the development effort for most projects, which is considerably higher than our observed frequency. This suggests that there needs to be more consideration for security aspects, especially amongst programming languages. Shell and PHP exhibit high rates of discussion (2.6\% and 2.1\%), approximately double the average rate, which indicates that security is properly appreciated for these languages. However, discussion and issues are almost non-existent for several languages, such as R, Coq, Julia, Fortran and MATLAB, which have an average rate of 0.1\%. This shows a lack of prioritisation or awareness of the security needs in development using these languages, which should be improved. 

Across the two sources, we find that the purpose of these discussions is most commonly caused by \textit{Errors} which users are trying to resolve. This indicates that secure development in most programming languages is difficult or error-prone. This is inverse to general development discussions \citep{beyer2020} which observe \textit{Errors} to be one of the less common question categories on Stack Overflow. We also observe C/C++ discussions to have a significantly higher focus on \textit{Review}, indicating a much higher examination of security for these languages. This aligns with the findings reported by \cite{bhattacharya2011} that code quality is higher for C/C++ software repositories. However, it has been found that there are a large number of software vulnerabilities in crowd-sourced code examples for C/C++ \citep{verdi2020, zhang2021}, which implies that security is not always the main consideration for this language. 

The majority of programming languages have Secure Development as the most dominant topic category. This shows that programming language specific challenges are more oriented towards development tasks, such as implementation, testing, and configuration; there is less regard for theory and concepts. These findings suggest that developers still need a better understanding and education of the security concepts that they are required to implement. However, we identify that the challenges of our created security taxonomy are prevalent across many different programming languages, which indicates that developers encounter a wide array of problems regardless of technology, and urges that more solutions need to be provided. 

From our analysis in Section 4, we identify that the security challenges and their characteristics are often oriented towards the task that the language is commonly utilized for. We focus on Web and Mobile development as many programming languages are explicitly applied for these domains \citep{sebesta2012}, such as JavaScript and TypeScript (Web), or Objective-C and Swift (Mobile). For general multi-purpose programming languages, such as Python and Go, it is more difficult to observe the trends. 

Web-Oriented languages have the most prevalent challenges in Access Control and Network Security, whereas Mobile-Oriented languages have the most prominent challenges in Cryptography, Access Control and Data Security. Another programming language trend we observed are for languages with a lower-level of abstraction. That is, languages which have strong coupling or control of the underlying operating system and architecture (i.e., C/C++, Shell). We observe that their posts on Stack Overflow receive the most popularity, as well as the most attention from expert users. 

We can speculate the reason for these trends through the application domain. For instance, if a programming language, such as JavaScript, is most commonly used for Web Development, then we can expect the major security issues to involve Access Control and Network Security. Web Development is a common task on GitHub, so we can also expect that these issues would receive high engagement (popularity) from the GitHub community. 

\subsection{Implications}
\textbf{\emph{For Developers.}} We firstly help practitioners to understand the impact that the choice of programming language can impose on secure development. Our findings can help developers to gain security awareness and insight into their selected programming languages. We identify what the prevalent security challenges of each language are, detail how common and readily the challenges are solved, how difficult the challenges are, and whether there is enough expertise in the community to help solve them. Through these findings, developers can identify what areas they should be better informed about and what areas they are able to get help in. We provide preliminary recommendations based on the most significant observations of our study. 

Firstly, developers should avoid security critical programming whilst using languages primarily designed for scientific programming (i.e., R, Julia and MATLAB). We observe in RQ1 that users of these languages have the lowest consideration for security, which implies that they are not designed for this task and receiving community support will be challenging. 

Developers should be more aware of the relation of security vulnerabilities to their specific development context, and how to defend against them. From our constructed security taxonomy, we found ``Software Vulnerabilities'' to be the least prominently discussed security topic for programming languages. Aside from ``Network Exploit'' discussions for web-oriented languages, and discussion of ``Memory Management'' for C/C++, these topics are largely non-existent for programming language focused discussion, which aligns with the findings of \cite{zahedi2018}. Whilst most vulnerabilities are language-independent, it is still critical that developers have specific implementation knowledge of how to avoid them. 

In RQ5 and RQ6, we observe that security implementation for TypeScript and JavaScript is relatively straightforward amongst the analysed communities. Challenges are resolved relatively quickly (low difficulty values on Stack Overflow), and issues receive a lot of attention (high popularity on GitHub). Despite the popularity and low difficulty, security users for these two languages are also generally of lower expertise than other languages. This could imply that sufficient security knowledge can be obtained relatively easily for these languages, as users are able to quickly share security knowledge and resolve issues, despite having lower average experience (as defined by GitHub expertise in Section 3.5).  

We also identify that the concerns and interests of developers vary between Stack Overflow and GitHub. For example, for RQ5 and RQ6 we find that Web-oriented languages have high popularity security issues on Stack Overflow but are low difficulty on GitHub, and that users have more security expertise for Mobile-oriented languages on GitHub than on Stack Overflow. These findings can help developers to find appropriate answers and discussions for their problems. We elaborate this further in Section 5.2. 

Additionally, our findings help motivate the creation of better security documentation and APIs for languages in the categories which face challenges. This effort will allow for easier secure use of the language and help reduce the faced challenges. For RQ2, we observe \textit{API Usage} to be the main reason for programming language security challenges on Stack Overflow. Furthermore, we find C/C++ to have a significantly higher proportion of discussions requesting for code review and improvement, implying that these users are often incapable of ensuring security on their own. Whilst users of C/C++ seem to be aware of the need for security, experienced developers should help ensure that appropriate resources and knowledge is available. 

We especially observe the need for better security support and resources for new programming languages, as we find users have significant security challenges after the initial release. This aligns with the findings of \cite{chakraborty2021}, who identify that adequate community resources are initially lacking for new programming languages. 

\textbf{\emph{For Managers.}} Although developers are responsible for their own personal programming language preferences, which may influence a project manager's decision, project managers ultimately have final say in programming language selection for a project \citep{naiditch1999}. Our study helps raise awareness of the importance of programming language selection and evaluation, promoting them to reserve project resources and time for these tasks. Through our comparison of security challenges for different languages we can help managers in their decision making for choosing what programming languages to use when starting a new project. Particularly through the characteristics of the challenges analysed in RQ5 and RQ6, we can obtain an indication of what languages are better suited to different categories of development. However, we acknowledge that not all languages can be used for all projects, and language is sometimes unable to be chosen due to external factors, such as legacy code. Thus, our findings and the identified challenges can also help guide necessary developer training decisions; either for new security development tasks, or for new programming language use. Additionally, educators can use the knowledge that we have synthesized from publicly available data as a potential guide for important topics in security education or a check-list for required security knowledge. 

\textbf{\emph{For Researchers.}} The challenges we have identified can help direct and motivate researchers to investigate potential solutions. The identified taxonomy of security issues can help direct and align this security language research. Furthermore, our proposed method of sampling, cleaning and analysing the data can be used by future researchers to identify the challenges and knowledge of other domains. The findings of our study also suggest that programming language should continue to be an important consideration in future security research, as we have observed the challenges and their nature to differ for different programming languages. Language consideration could also be potentially strengthened in several research areas. For instance, vulnerability prediction models often do not consider the characteristics or features of the languages that they target \citep{vuldigger,croft2021empirical}. 

Additionally, we arouse the attention of language designers for the impacts that programming language can have on secure development. Our findings help identify the major security issues and considerations of developers, which can assist designers in improving the supportive capabilities of security features in programming languages. 

\subsection{Difference in Discussion between Stack Overflow and GitHub }
Interestingly, for RQ3-6 we observe a consistent disconnect between the security characteristics of programming languages on Stack Overflow and GitHub. For instance, PHP and Perl are some of the most difficult languages for security on GitHub but the least difficult on Stack Overflow. Furthermore, the highest popularity security issues on GitHub are for TypeScript and JavaScript, whereas the issues for these languages are relatively unpopular on Stack Overflow. This separation is consistent with the analysis reported by \cite{han2020}, who also observe a disparity between Stack Overflow and GitHub. 

Through this disconnect, we can derive recommendations for where users should go for assistance with security related challenges of a particular language. Although users do not traditionally visit GitHub for answers to programming problems, GitHub issues of open source projects are still a viable resource for developers to query project specific questions; questions have even been observed to be the third most common issue type \cite{cabot2015exploring}.  Hence, for languages more oriented towards software and app development (e.g., JavaScript, TypeScript, Swift, Objective-C), users should seek answers on GitHub, as these challenges are more popular, less difficult and answered by more experienced users on this platform. GitHub also exhibits a higher general rate of discussion for Secure Development. For other languages, particularly general-purpose languages (e.g., Java, Python) or languages with lower abstraction (e.g., C/C++, Shell), users should seek help from the Stack Overflow community. 

We conjecture that work should be done in connecting these two sources, as knowledge gained from one source may not adequately prepare users. This is showcased through user expertise, where we observe that Stack Overflow users are relatively inexperienced with Mobile-Oriented languages, whereas the users of these languages on GitHub exhibit the most experience. 

\subsection{Vulnerabilities for Different Programming Languages}
Another approach for determining the security of a programming language is to examine how prone it is to vulnerabilities. By adopting a similar approach to \cite{decan2018} and \cite{gkortzis2018}, we can determine an indication of language vulnerabilities in their packages and projects, via the vulnerability advisories \emph{Snyk.io} and \emph{National Vulnerability Database (NVD)}, respectively. Through this method we identify that projects written in C and PHP, and packages written for Java, have the highest number of reported vulnerabilities. However, few assertions can be made from this information. It does not take into account the prevalence of the language; C is one of the oldest popular programming languages, and Java is one of the most commonly used languages. Similarly, vulnerabilities in certain languages are more well known and easier to detect \citep{shahriar2012}, and the number of annual reported vulnerabilities is generally increasing for all languages\footnote{\url{https://www.whitesourcesoftware.com/most-secure-programming-languages/}}.

Furthermore, the representation and reliability of NVD data has also been questioned by several previous works \citep{massacci2010, nguyen2013}. NVD's nature of only reporting observed and documented vulnerabilities has been shown to potentially obscure and misconstrue the true vulnerabilities of a project \citep{massacci2010}. \cite{nguyen2013} also investigated the reported vulnerabilities in Google Chrome and found several errors. It is further observed that developers themselves do not necessarily perform consistent reporting of vulnerabilities. For instance, this GitHub issue\footnote{\url{https://github.com/bcit-ci/CodeIgniter/issues/4020}} discusses whether certain vulnerabilities should be reported to a vulnerability database.  Thus, vulnerability databases cannot be considered as a definitive view of software vulnerabilities.  

For the purposes of this study, we do not explicitly consider vulnerabilities as it is difficult to draw conclusions from this data alone; vulnerabilities are not directly equivalent to secure development challenges. Vulnerabilities only indicate the specific exploits and weaknesses in a system. They also rely on proper identification, documentation and reporting. It might be expected that some developers are unwilling to document or discuss vulnerabilities in their software publicly. Instead we try to examine the complete picture by also investigating secure implementation, coding, defenses, knowledge and theory. 

\subsection{Reliability of Crowd-sourced Knowledge}
Despite the existence of crowd-sourced security knowledge in our examined data sources, its reliability has been a source of contention in past research. There is evidence that code snippets on Stack Overflow contain security flaws \citep{acar2016,fischer2017}, and a recent study into the security answer posts on Stack Overflow found almost 45\% to be insecure \citep{chen2019}. However, approximately 70\% of questions on Stack Overflow receive an accepted answer\footnote{\url{https://stackoverflow.com/questions?tab=Unanswered}}, which implies that discussion and knowledge-sharing is often successful and informative for the original poster in the least. 

The success and validity of security discussion in GitHub issues also lacks confirmation. We find that as of July 2020, 88\% of the issues in our GitHub dataset are closed. If we similarly consider closed GitHub issues as a discussion with a successful resolution, then this also implies that the majority of GitHub issue discussions achieve a sufficient level of conclusion. 

\section{Threats to Validity}
\textbf{\emph{Internal Validity. }}
We use topic modelling to automatically cluster documents, and identify topics of discussion. Due to the size of our dataset, we are required to use the semi-automated approach of topic modelling for analysis. However, the topics and clusters produced by topic modelling are often volatile \citep{mantyla2018}, and the interpretation of the produced topics is often subjective. We note that the topics we identify in Section 4.3. are not exhaustive or fully representative of the security challenges, but are the most prevalent in the text data. We use manual validation to help confirm the accuracy of the topics, and cross-check our labels to help confirm their validity. Furthermore, although the metrics outlined in Section 3.5 are able to give an indication of the characteristics of the data, we acknowledge that they are not perfect.

The rigidness of the produced topics also limits the versatility of our identified security categories and challenges. Documents can only be assigned to the available topics. If a security topic is not reported for a language, it often does not mean that it does not exist, but that is obscured by a more dominant topic. 

Due to the nature of our empirical study, we acknowledge that there are threats which affect our conclusions regarding the correlation between use of programming language and security challenges. Our findings may be influenced by several confounding factors, such as expertise of the users behind the discussions. However, we observe several potentially interesting differences across the security challenges, which we report. These can be used as a starting point for future research to help confirm our findings. 

Similarly, the large size scale of the analysis prevents us from providing specific insight into the experience or engagement within these programming language communities, and we hence use popularity, difficulty and expertise metrics defined in Section 3.5 as proxies for our analysis. However, we are still able to provide important insight into the prevalent security activity within these different communities. 

\textbf{\emph{External Validity. }}
We conduct an extremely large-scale study to help improve the generalizability of our findings by investigating two of the most prominent sources for developer community support; Stack Overflow and GitHub. However, Stack Overflow and GitHub may not be perfectly representative of all developers. We also discuss in Section 5.2 and Section 5.3 that our findings are likely to change across different datasets.  Due to the nature of our study and methods, our findings can only be considered reliable for our selected datasets. 

For GitHub, our method of data collection has further flaws due to our repository sampling method. The data of each language may be biased towards the repositories that were sampled, which may obscure the true trends of the underlying language. However, this assertion can be correct for several other similar empirical studies. 

As the data sources from which we extract the data from are subject to change, our study is not perfectly reproducible. For both Stack Overflow and GitHub, posts can be added, removed, deleted, changed or updated, and thus someone using the same method as us would likely extract a slightly different dataset. Similarly, the posts referenced in this paper for discussion may not always be available if they are deleted or removed. 

\textbf{\emph{Construct Validity. }}
Keyword and tag matching are not perfect methods for data extraction, as they rely on a manually constructed keyword set and correct use of the keywords by the users of the dataset (e.g., correct spelling, appropriate post tagging). However, we manually review several samples of posts in our dataset to help confirm their relevance to the topic our study, and thus help confirm the validity of our approach. 

There is also the threat to validity regarding the implementation of the study. We independently reviewed our scripts and code to check for correctness and reduce this potential threat. We have also made our data and results publicly available for reproducibility. 

\section{Conclusion and Future Work}
We have presented a large-scale comparative analysis of the security challenges encountered during development using different programming languages. To conduct this analysis, we examined approximately 280,000 publicly accessible security-related developer discussions from Stack Overflow and GitHub for 15 popular programming languages. We also investigated the nature of these challenges and their ability to be resolved to help provide recommendations to developers. 

Our findings show that languages exhibit varying rates of security discussion, categories of security issues, and characteristics in which these security issues are handled by the community. Language security trends emerge for languages with similarly oriented domains, particularly web-oriented languages, mobile-oriented languages, and lower-level abstraction languages. The observed security challenges are handled better by different languages for different categories and different sources. For example, it is suggested that TypeScript is well suited for ``Network Security'' and avoiding ``Network Exploits'', as we observe the discussed issues to have high popularity and low difficulty amongst the community.

Our study is expected to motivate the need for practitioners to consider and evaluate the influence that programming languages might impose on secure software implementation. Our findings can help them understand and navigate the secure development landscape for different languages. Additionally, our identified taxonomy of security challenges have the potential to guide and increase the productivity of developer training, or secure software engineering research. 

For future work we plan to investigate whether the difference in challenges for different programming languages is also reflected in the code and software. To achieve this we intend to analyse the explicit security situations of different open-source repositories in more detail, by examining the vulnerabilities, software metrics, and code quality. We particularly aim to investigate Software Vulnerability related issues, as we noticed discussion for these topics to be lacking for most programming languages. Additionally, we intend to extend our study to also consider technology stacks (i.e., libraries and frameworks). We also aim to conduct more fine-grained analysis into users' expertise and engagement through more extensive qualitative analysis or a user study. This additional analysis should help support our findings and provide more specific insights. Furthermore, our analysis in Fig. \ref{fig:cat_comparison} reveals insights into how open source security challenges align and intersect with more formal security taxonomies from CyBoK and CWE. We aim to investigate in future how these software and security communities can be brought together. 

\section{Acknowledgements}
We would also like to sincerely thank Peter Sestoft and Triet Le, as well as the anonymous reviewers for the insightful and constructive comments they provided towards improving the paper. The work has been supported by the Cyber Security Research Centre Limited whose activities are partially funded by the Australian Government’s Cooperative Research Centres Programme. The work was also supported with super-computing resources provided by the Phoenix High Powered Computing (HPC) service at the University of Adelaide. 

\section{Declarations}
\section*{Funding}
The work has been supported by the Cyber Security Research Centre Limited whose activities are partially funded by the Australian Government’s Cooperative Research Centres Programme. 
\section*{Conflicts of interest}
Not applicable. 
\section*{Availability of data}
All data has been made available via an online appendix. 
\section*{Code availability}
Not applicable.

\bibliographystyle{spbasic}
\bibliography{bibfile.bib}

\end{document}